\documentclass[12pt]{article}

\usepackage[margin=1in]{geometry}
\usepackage{amsmath,amssymb}
\usepackage{siunitx}
\usepackage{graphicx}
\usepackage{booktabs}
\usepackage{enumitem}
\usepackage{authblk}
\usepackage{multirow}
\usepackage{float}
\usepackage{xcolor}
\usepackage{cite}
\usepackage{hyperref}
\usepackage[capitalise]{cleveref}
\usepackage{subcaption}

\graphicspath{{./}}

\newcommand{\todo}[1]{\textcolor{red}{#1}}

\hypersetup{
  colorlinks=true,
  linkcolor=black,
  citecolor=black,
  urlcolor=blue
}

\title{\textbf{Transferable Graph Neural Network Surrogates for Molecular Dynamics Across Crystal Symmetries}}

\author[1]{Judah Immanuel}
\author[2]{Avik Mahata\thanks{Corresponding author: \textit{mahataa@merrimack.edu}}}
\author[3]{Aniruddha Maiti}

\affil[1]{Department of Computer Science, Merrimack College, North Andover, MA, USA}
\affil[2]{Department of Mechanical and Electrical Engineering, Merrimack College, North Andover, MA, USA}
\affil[3]{Department of Mathematics, Engineering, and Computer Science,\\ West Virginia State University, Institute, WV, USA}

\date{}

\begin{document}
\maketitle

\begin{abstract}
We present a transferable graph neural network (GNN) surrogate 
framework for molecular dynamics (MD) that directly predicts atomic 
displacements and propagates atomistic configurations without explicit 
force evaluation or numerical time integration. The central objective 
of this work is to establish whether a common GNN formulation can 
represent atomic dynamics across materials with fundamentally different 
crystal symmetries and coordination environments. The same network 
architecture, feature representation, graph construction, and training 
protocol are applied without symmetry-specific modification to 
face-centered cubic (FCC) aluminum, body-centered cubic (BCC) iron, 
and hexagonal close-packed (HCP) magnesium. Across these distinct 
elemental and crystallographic systems, the framework achieves 
position-prediction errors on the order of $10^{-4}\,\text{\AA}^2$ 
and supports stable autoregressive propagation to nanosecond time 
scales. The predicted trajectories preserve thermodynamic stability, 
characteristic coordination-shell structure in the radial distribution 
functions, and temperature-dependent mean-squared-displacement behavior. 
In particular, the framework captures the closely spaced coordination 
shells of BCC iron and the anisotropic coordination environment of HCP 
magnesium without introducing lattice-specific representations. These 
results demonstrate that direct GNN-based atomic propagation can be 
formulated as a transferable framework across different elements, 
lattice symmetries, and coordination geometries, providing a pathway 
toward generalizable surrogate models for accelerated molecular dynamics.
\end{abstract}

\section{Introduction}

Molecular dynamics is a deterministic framework grounded in classical statistical mechanics. By integrating Newton's equations of motion for a system of interacting atoms, it provides a direct connection between microscopic trajectories in phase space and emergent macroscopic behavior. This foundation has made molecular dynamics a cornerstone of physics, chemistry, and materials engineering \cite{allen_tildesley, frenkel_smit, rapaport}. In metals and alloys it has been especially productive, clarifying solidification pathways, defect nucleation, plastic deformation, diffusion, and fracture in ways that complement both experiment and continuum modeling \cite{mishin1999interatomic, bitzek_deformation, marian2004dynamic}. The practical reach of the method is nevertheless bounded by a tradeoff between accuracy and cost. Ab initio molecular dynamics based on density functional theory offers a accurate description of interatomic interactions \cite{hohenberg1964, kohn1965, marx_hutter, kresse_furthmuller}, but its expense confines simulations to small cells and short times. Classical molecular dynamics avoids that expense by adopting empirical potentials such as the embedded atom method \cite{daw1984eam, finnis1984}, which allows millions of atoms to be simulated at the cost of reduced transferability across compositions and structural environments. A second and more fundamental constraint applies regardless of how the forces are obtained. Numerical stability requires that the integration step resolve the fastest atomic vibrations, which fixes the step at the femtosecond scale. Reaching experimentally relevant times therefore demands millions of steps, and the cost becomes prohibitive even for simple metals \cite{kadau2006molecular, voter2002extending, perez2009accelerated}. The computational burden becomes even more severe for processes such as 
solidification, defect evolution, diffusion, and structural transformations, 
where both large system sizes and extended trajectories are often required to 
resolve the relevant mechanisms and statistics \cite{Mahata_2018, MAHATA2019125255, MAHATA2022110902, MAHATA2026114443}. Moreover, improving the 
physical fidelity of the interatomic description generally increases the cost 
of each force evaluation, further amplifying the expense of conventional 
time-stepping. Thus, even as increasingly accurate interatomic potentials 
improve the description of complex atomic environments, the need to evaluate 
forces and advance the system through millions of sequential integration steps 
remains a fundamental computational bottleneck. This motivates surrogate 
approaches that target not only the cost of evaluating the interatomic 
interactions, but the time-propagation procedure itself.

Machine learning entered this field first through the accuracy channel. Beginning with high-dimensional neural network potentials \cite{behler2007generalized} and descriptor-based regression \cite{bartok2013soap}, the field has advanced rapidly toward graph-based architectures including SchNet \cite{schuett2017schnet}, DeePMD \cite{wang2018deepmd}, NequIP \cite{batzner2022nequip}, MACE \cite{batatia2022mace}, and Allegro \cite{musaelian2023allegro}, together with universal models trained across the periodic table \cite{chen2022m3gnet, deng2023chgnet, merchant2023gnome, batatia2023foundation}. These machine-learned interatomic potentials now approach the accuracy of density functional theory for energies and forces at a small fraction of the cost \cite{unke2021review, zuo2020performance}, and they have been scaled to systems of one hundred million atoms \cite{jia2020pushing}. What they do not do is change the structure of the simulation. The integration loop remains, the step remains femtosecond in size, and the number of steps required to reach a nanosecond remains of order one million. Even in the limit of a free force evaluation, the timescale barrier stands. A different and more ambitious response is to replace the integration loop itself with a learned operator. Rather than predicting forces to be integrated, such a model maps the system state at time $t$ to its state at time $t + \Delta t$ in a single evaluation, with $\Delta t$ chosen far larger than any stable integration step. Graph network simulators established that this is feasible for particle systems while preserving permutation invariance and locality \cite{sanchez2020learning, pfaff2021meshgraphnets, battaglia2018relational}, and related efforts have explored time-coarsened dynamics \cite{klein2023timewarp}, conditional diffusion propagators \cite{hsu2024score}, generative trajectory models \cite{jing2024mdgen}, and learned effective dynamics \cite{vlachas2022multiscale}. The obstacles that have limited this line of work are well documented. Autoregressive rollouts accumulate error, conserved quantities drift, and structural fidelity degrades over long horizons \cite{fu2023forces, stocker2022robust, bihani2023stability}. Most demonstrations have consequently been restricted to a single chemistry, a single crystal structure, or a narrow window of thermodynamic conditions. We previously addressed the first of these obstacles by introducing a transformer graph neural simulator that predicts atomic displacements directly and remains stable over long rollouts, and we validated it on face-centered cubic aluminum \todo{\cite{mahata2025gnnal}}. That work left open the question that motivates the present study. A direct propagator learns the evolution operator of a specific lattice through the local geometry it observes during training. Whether the same architecture can learn a different operator for a different lattice, without symmetry-adapted bases, structure-specific descriptors, or any architectural change, is not answered by success on a single close-packed cubic metal. The concern is concrete rather than abstract. Face-centered cubic aluminum has twelve nearest neighbors at a single well separated distance and an isotropic elastic response. Body-centered cubic iron has eight nearest neighbors at 2.48~\AA{} followed immediately by six second neighbors at 2.87~\AA, so the first two coordination shells overlap and the model must resolve them from a distance encoding alone. Hexagonal close-packed magnesium is not cubic at all. Its coordination shell splits into six in-plane and six out-of-plane neighbors whose separations differ through the axial ratio, and its vibrational and diffusive response is anisotropic by construction \cite{agnew2005plastic, barnett2007twinning}. If a single set of learned parameters can be fit to each of these three environments using an identical pipeline, then the direct propagator is a property of the method rather than an artifact of a favorable test case.

To examine the transferability of the framework across distinct crystal 
symmetries and coordination environments, we apply it to BCC iron and HCP magnesium, while 
retaining the same architecture, feature set, graph construction, and 
training protocol used for FCC aluminum. These 
systems provide contrasting structural environments for evaluating the 
generality of the approach, ranging from the cubic coordination of BCC 
iron to the anisotropic local geometry of HCP magnesium. They are also 
of broad technological relevance. Iron forms the basis of steels, where 
high-temperature behavior is important to radiation damage evolution 
and phase stability in structural alloys 
\cite{malerba2010comparison,dudarev2013density,gilbert2008structure}, 
while magnesium is the lightest structural metal in engineering use, 
with its limited room-temperature ductility closely associated with the 
anisotropy of HCP deformation 
\cite{pollock2010weight,wu2018mechanistic}. The resulting surrogate 
models are evaluated through displacement accuracy against reference 
trajectories, kinetic-energy stability during autoregressive rollouts 
extending to one nanosecond, structural fidelity through radial 
distribution functions, and dynamical consistency through 
temperature-dependent mean squared displacement. Finally, ablation 
studies are used to assess the contributions of the structural 
descriptors and model capacity across the different coordination 
environments.onsistent with their differing coordination geometry.

\section{Methodology}

\subsection{Molecular Dynamics}

Reference trajectories for both metals were generated with LAMMPS \cite{plimpton1995,thompson2022lammps}. Iron was modeled in the body-centered cubic structure with a lattice parameter of 2.855~\AA{} using the embedded atom method (EAM) potential developed by Mendelev \emph{et al.} \cite{Mendelev2003}. Magnesium was modeled in the hexagonal close-packed structure with lattice parameters $a = 3.19$~\AA{} and $c = 5.17$~\AA{} using the EAM potential of Zhou \emph{et al.} \cite{Zhou2004}. Each simulation employed periodic boundary conditions in all three spatial directions. The supercells contained 2000 atoms for bcc Fe and 3600 atoms for hcp Mg. Systems were equilibrated for 50~ps in the isothermal isobaric ensemble at zero pressure and then propagated in the canonical ensemble with a Nos\'{e} and Hoover thermostat \cite{nose1984, hoover1985}. The equations of motion were integrated using the velocity Verlet algorithm \cite{verlet1967,swope1982} with a timestep of 1~fs. Atomic trajectories were written at every integration timestep (1~fs), ensuring that all configurations generated during the simulations were retained for subsequent analysis and graph construction, rather than using temporally subsampled trajectories. Iron was sampled at 400, 800, 900, 1000, 1100, and 1300~K, which spans the ferritic regime and extends well above the temperatures at which anharmonic effects become pronounced. Magnesium was sampled at 300, 500, and 800~K. The set of trajectory lengths used for each evaluation is summarized in \cref{tab:sim_conditions}. Per-atom structural descriptors were computed with OVITO \cite{stukowski2010ovito} and comprise the centrosymmetry parameter \cite{kelchner1998csp}, the common neighbor analysis label \cite{honeycutt1987cna}, the first two Minkowski structure metrics \cite{steinhardt1983bond, mickel2013minkowski, lechner2008accurate}, and the coordination number.

\begin{table}[H]
\centering
\caption{Reference molecular dynamics conditions used for training and evaluation. All trajectories were generated with periodic boundary conditions, a timestep of 1~fs, and configurations were recorded every timestep.}
\label{tab:sim_conditions}
\begin{tabular}{llll}
\toprule
\textbf{System} & \textbf{Structure} & \textbf{Temperature range (K)} & \textbf{Trajectory length} \\
\midrule
Iron       & Body-centered cubic      & 300--1300 (100 K intervals) & 25 ps \\
Magnesium  & Hexagonal close-packed   & 300--800 (100 K intervals)  & 25 ps \\
\bottomrule
\end{tabular}
\end{table}

\subsection{Graph Representation and Model Architecture}

The architecture is unchanged from our previous work on face-centered cubic aluminum \todo{\cite{mahata2025gnnal}}, and we summarize it here only to the extent needed to make the present results reproducible. Full derivations and design ablations are given in previously published work \cite{mahata2025gnnal}.

\paragraph{Graph construction.} Each configuration is represented as a graph $G = (V, E)$ in which nodes are atoms and edges connect pairs of atoms separated by less than a cutoff of $6.287$~\AA. The graph is rebuilt independently at every frame from the current positions, so the model performs per-frame inference and carries no hidden state between frames. Minimum image conventions are applied to all relative vectors so that periodicity is respected.

\paragraph{Features.} Each node carries its Cartesian position, its velocity, a learnable embedding of the atomic number, radial basis expansions of the prediction horizon $\Delta t$ and the temperature $T$, and a learned projection of the five structural descriptors listed above. Each edge carries a radial basis expansion of the interatomic distance together with a learned transformation of the relative velocity. Radial basis expansions use $N_\mathrm{rbf} = 32$ centers spaced linearly over the relevant input range,
\begin{equation}
\psi_\mathrm{rbf}(s)_k = \exp\!\left[-\gamma\,(s - c_k)^2\right], \qquad k = 1,\ldots,N_\mathrm{rbf},
\end{equation}
with $\gamma = 1/\Delta_c^2$ where $\Delta_c$ is the spacing between adjacent centers. Because $\Delta t$ and $T$ enter as node features, a single trained model serves the entire temperature range and the entire set of prediction horizons rather than requiring one model per condition. Feature dimensions are collected in \cref{tab:node_edge_features}.

\begin{table}[H]
\centering
\caption{Node and edge features supplied to the network. The same feature set is used for iron and for magnesium without modification.}
\label{tab:node_edge_features}
\begin{tabular}{l l c}
\toprule
\multicolumn{3}{c}{\textbf{Node features}} \\
\midrule
\textbf{Feature} & \textbf{Transformation} & \textbf{Dimension} \\
\midrule
Atomic position $(x_i, y_i, z_i)$ & Raw coordinates & 3 \\
Atomic velocity $(v_{x,i}, v_{y,i}, v_{z,i})$ & Raw components & 3 \\
Atomic number $z_i$ & Trainable embedding & 64 \\
Prediction horizon $\Delta t$ & Radial basis expansion & 32 \\
Temperature $T$ & Radial basis expansion & 32 \\
\midrule
Centrosymmetry parameter & \multirow{5}{*}{Shared linear layer with GELU} & \multirow{5}{*}{64 (total)} \\
Common neighbor analysis &  &  \\
Minkowski structure metric $q_1$ &  &  \\
Minkowski structure metric $q_2$ &  &  \\
Coordination number &  &  \\
\midrule
\textbf{Total} & & \textbf{198} \\
\midrule
\multicolumn{3}{c}{\textbf{Edge features}} \\
\midrule
\textbf{Feature} & \textbf{Transformation} & \textbf{Dimension} \\
\midrule
Interatomic distance $\lVert \mathbf{r}_j - \mathbf{r}_i \rVert$ & Radial basis expansion & 32 \\
Relative velocity $\mathbf{v}_j - \mathbf{v}_i$ & Two-layer perceptron with GELU & 64 \\
\midrule
\textbf{Total} & & \textbf{96} \\
\bottomrule
\end{tabular}
\end{table}

\paragraph{Message passing and attention.} The network consists of $N = 6$ identical blocks. Within each block, node states are first updated by message passing \cite{gilmer2017mpnn},
\begin{equation}
\mathbf{m}_{ij} = \phi_m(\mathbf{h}_i, \mathbf{h}_j, \mathbf{e}_{ij}), \qquad
\mathbf{a}_i = \frac{1}{|\mathcal{N}(i)|} \sum_{j \in \mathcal{N}(i)} \mathbf{m}_{ij}, \qquad
\mathbf{h}_i' = \phi_u(\mathbf{h}_i, \mathbf{a}_i),
\end{equation}
and then refined by graph self-attention \cite{vaswani2017attention, velickovic2018gat, shi2021masked}. Queries are formed from the central node while keys and values combine neighbor states with edge features,
\begin{equation}
\mathbf{q}_i = W_q \mathbf{h}_i, \qquad
\mathbf{k}_{ij} = W_k^h \mathbf{h}_j + W_k^e \mathbf{e}_{ij}, \qquad
\mathbf{v}_{ij} = W_v^h \mathbf{h}_j + W_v^e \mathbf{e}_{ij},
\end{equation}
\begin{equation}
\alpha_{ij} = \frac{\exp\!\left(\mathbf{q}_i^{\top} \mathbf{k}_{ij} / \sqrt{d}\right)}{\sum_{k \in \mathcal{N}(i)} \exp\!\left(\mathbf{q}_i^{\top} \mathbf{k}_{ik} / \sqrt{d}\right)}, \qquad
\mathbf{m}_i = \sum_{j \in \mathcal{N}(i)} \alpha_{ij} \mathbf{v}_{ij}.
\end{equation}
Residual connections and layer normalization \cite{ba2016layernorm} surround both the attention operation and a subsequent feedforward network with GELU activations \cite{hendrycks2016gelu}. Restricting attention to graph edges rather than to all atom pairs is what allows the model to weight neighbors by learned relevance while retaining the locality that makes the cost linear in system size. This matters for the present study, because the relevance of a neighbor is precisely what differs between a body-centered cubic environment, in which the first and second shells nearly merge, and a hexagonal close-packed environment, in which basal and non-basal neighbors are geometrically inequivalent.

\paragraph{Iterative position update.} Each block emits a displacement that is applied immediately,
\begin{equation}
\Delta \mathbf{r}_i^{(b)} = W_\mathrm{out}\,\mathbf{h}_i^{\mathrm{attn},(b)}, \qquad
\mathbf{r}_i \leftarrow \mathbf{r}_i + \Delta \mathbf{r}_i^{(b)},
\end{equation}
so that the final prediction is the accumulation over all blocks,
\begin{equation}
\mathbf{r}_i^{\,t+\Delta t} = \mathbf{r}_i^{\,t} + \sum_{b=1}^{N} \Delta \mathbf{r}_i^{(b)}.
\end{equation}
Updating positions inside the forward pass rather than only at the output layer allows later blocks to observe partially corrected geometry, which turns a single large displacement prediction into a sequence of smaller refinements. No force is ever computed, and no numerical integration is performed.

\subsection{Training}

Models were implemented in PyTorch \cite{paszke2019pytorch} with 
PyTorch Geometric \cite{fey2019pyg} and trained separately for iron 
and magnesium using identical hyperparameters. The objective function 
was the mean squared error (MSE) between the predicted and reference 
atomic positions,
\begin{equation}
\mathcal{L} =
\frac{1}{N_\mathrm{atoms}}
\sum_{i=1}^{N_\mathrm{atoms}}
\left\lVert
\hat{\mathbf{r}}_i^{\,t+\Delta t}
-
\mathbf{r}_i^{\,t+\Delta t}
\right\rVert^2 .
\end{equation}

The network parameters were optimized using AdamW
\cite{loshchilov2019adamw}. Frames were partitioned into training and
validation sets using an 80:20 temporal split for each temperature,
such that both partitions contained configurations from every
temperature considered.

To improve stability during multistep propagation, training employed a
staged curriculum in the autoregressive rollout length
\cite{bengio2009curriculum}. Training began with a single-step rollout
for the first five epochs. The rollout length was subsequently
increased to two, three, and four steps at epochs 5, 15, and 25,
respectively. At each rollout step, the predicted atomic configuration
was used as input to the subsequent step, allowing prediction errors to
propagate through the training trajectory. This curriculum progressively
exposes the model to longer autoregressive trajectories, enabling it to
first learn short-horizon atomic evolution before optimizing against
errors accumulated during multistep propagation.

\subsection{Evaluation}

Trained models were evaluated by autoregressive rollout. The predicted configuration at each step is used as the input for the subsequent prediction. In this scheme, errors are free to compound, and any instability in the learned operator can therefore be directly observed through trajectory drift or structural collapse. Rollouts were carried out to the lengths listed in \cref{tab:sim_conditions} and were analyzed with three physical observables.

The kinetic energy of the cell reports on thermodynamic consistency. We plot the deviation of the instantaneous kinetic energy from its reference value, smoothed over a 50-step window, and look for the absence of secular drift rather than for exact instantaneous agreement, since a thermostatted trajectory is expected to fluctuate.

The radial distribution function reports on structural fidelity,
\begin{equation}
g(r) = \frac{1}{4\pi r^2 \rho N} \sum_{i} \sum_{j \neq i} \left\langle \delta(r - r_{ij}) \right\rangle ,
\end{equation}
where $\rho$ is the number density. Because $g(r)$ resolves individual coordination shells, it is a direct test of whether the surrogate has learned the specific lattice geometry rather than a generic condensed phase.

The mean squared displacement reports on dynamical fidelity,
\begin{equation}
\mathrm{MSD}(t) = \frac{1}{N} \sum_{i=1}^{N} \left\lVert \mathbf{r}_i(t) - \mathbf{r}_i(0) \right\rVert^2 .
\end{equation}
In a crystalline solid below the melting point the mean squared displacement saturates at a plateau set by the vibrational amplitude, and the height of that plateau increases monotonically with temperature. A surrogate that has learned only to keep atoms near their sites will produce a plateau, but only one that has learned the correct temperature-dependent anharmonicity will produce plateaus at the correct heights and in the correct order.

\section{Results and Discussion}

\subsection{Training Behavior and Displacement Accuracy}

\subsubsection{Training Convergence and Displacement Accuracy in BCC Iron}

\begin{figure}[H]
    \centering
    \includegraphics[width=1.0\linewidth]{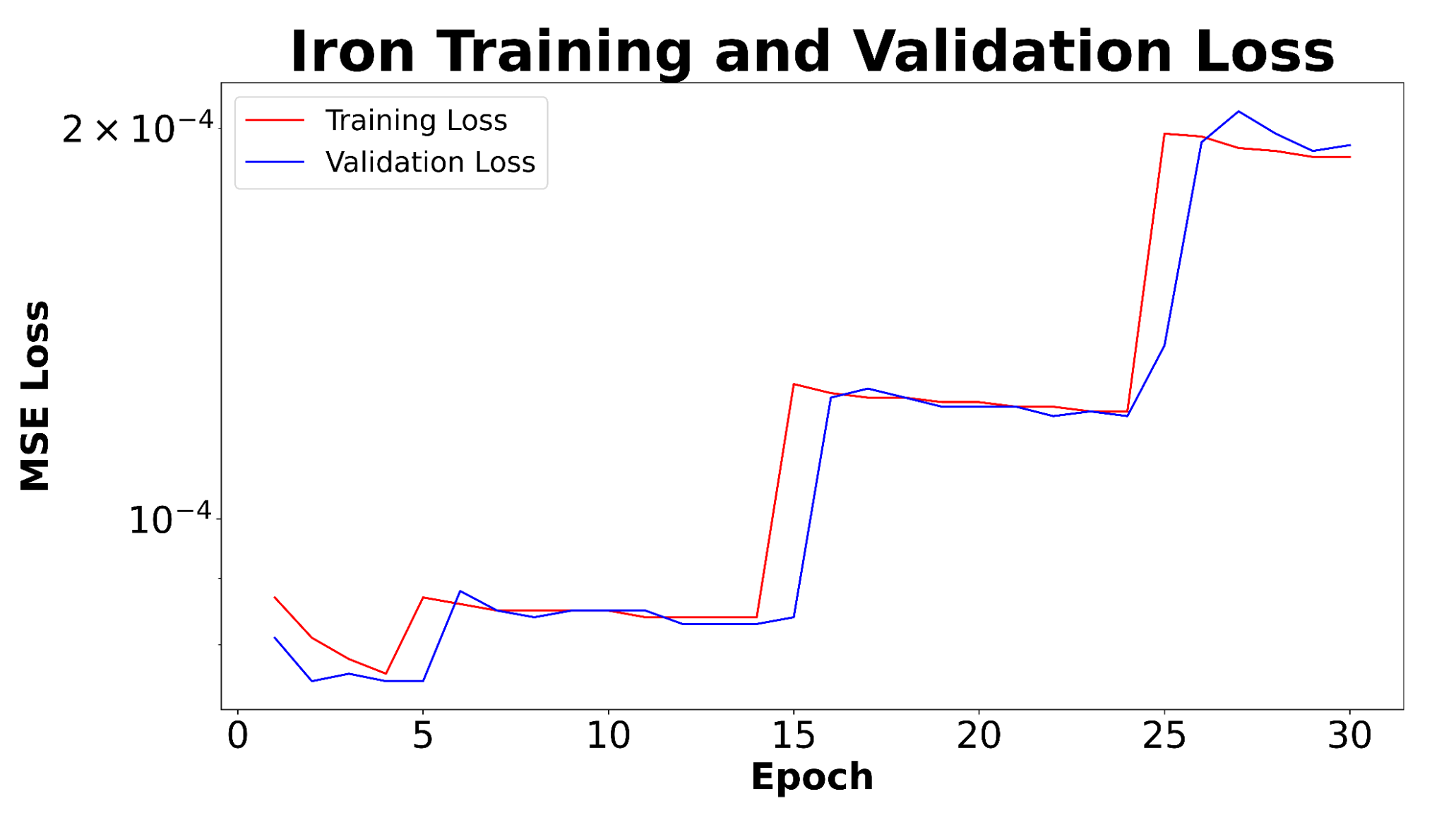}
    \caption{Training and validation loss for the iron model over 30 epochs. 
    The mean squared error is reported in units of \AA$^2$ on a logarithmic 
    scale. The three upward steps at epochs 5, 15, and 25 correspond to 
    increases in the autoregressive rollout length during curriculum training. 
    Following each transition, the loss re-converges within a small number of 
    epochs. The close agreement between the training and validation curves 
    indicates that the model does not exhibit significant overfitting.}
    \label{fig:fe-loss}
\end{figure}

\cref{fig:fe-loss} shows the training and validation MSE loss at different epochs for iron. The loss falls to approximately $7 \times 10^{-5}$~\AA$^2$ within four epochs. At epoch 5, the autoregressive rollout horizon is increased, and a sharp increase in the loss is observed. The loss then gradually decreases until epoch 15, when the rollout horizon is increased again and another distinct increase in the loss is observed. The same behavior occurs at epoch 25. Following each curriculum transition, the loss progressively decreases as training continues with the longer rollout horizon.
Because the difficulty of the regression grows with the horizon, an increase in the loss at these transitions is expected and does not indicate a failure of optimization. The relevant observations are that after each increase in the rollout horizon, the loss decreases within the following epochs, which shows that the model continues to improve under the more difficult prediction task. At the end of training, the loss reaches approximately $1.9 \times 10^{-4}$~\AA$^2$. Taking the square root gives a typical per-atom position error of roughly $1.4 \times 10^{-2}$~\AA{} at the longest horizon, which is more than an order of magnitude smaller than the thermal vibrational amplitude at these temperatures and two orders of magnitude smaller than the nearest neighbor separation of 2.48~\AA. As the training and validation curves overlap closely throughout, including immediately after each curriculum transition, it can be concluded that the model generalizes to held-out frames at every horizon rather than memorizing the training trajectories.

\subsubsection{Training Convergence and Displacement Accuracy in HCP Magnesium}

\begin{figure}[H]
    \centering
    \includegraphics[width=1.0\linewidth]{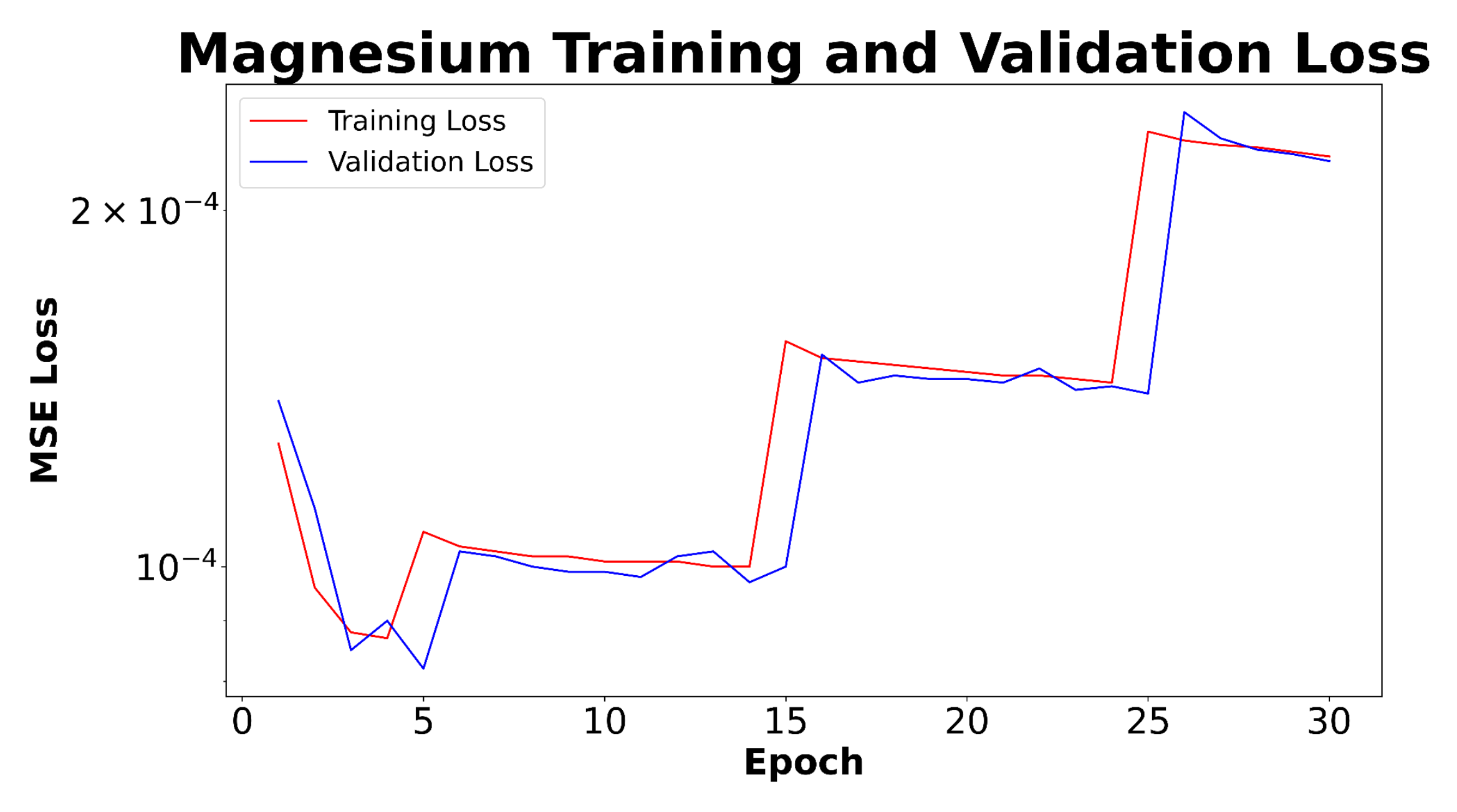}
    \caption{Training and validation loss for the magnesium model over 30 epochs. The optimization history follows the same qualitative pattern as iron, with a rapid initial decrease, a stable plateau, and two curriculum transitions at epochs 15 and 25. The larger fluctuations in the early epochs reflect the anisotropy of the hexagonal close-packed environment.}
    \label{fig:mg-loss}
\end{figure}

As shown in \cref{fig:mg-loss}, a similar trend is observed for magnesium. During initial epochs, the loss decreases sharply  to $7.5 \times 10^{-5}$~\AA$^2$. Distinct increases in MSE loss are observed at epochs 5, 15, and 25, when the autoregressive rollout horizon is increased. After each transition, both the training and validation losses decrease again as optimization continues with the longer rollout horizon. It can be observed, the early epochs are noisier than for iron. We attribute this to the anisotropy of the hexagonal close-packed environment. In-plane and out-of-plane neighbors sit at different distances and support different vibrational amplitudes, so the network must resolve two inequivalent local geometries before its representation stabilizes, whereas the body-centered cubic environment of iron is isotropic at the level of the first two shells taken together. From \cref{fig:fe-loss} and \cref{fig:mg-loss}, it should be noted that similar trends in loss are observed for both iron and magnesium. In both cases, the training and validation losses decrease between successive curriculum transitions. These results show that the same GNN architecture and training approach can be applied to both BCC iron and HCP magnesium without a symmetry-specific modification of the model

\subsection{Thermodynamic Consistency}
Stability under autoregressive rollout provides an important measure of the model's ability to reproduce atomic trajectories beyond single-step predictions. If the learned evolution introduces a systematic energy error at each prediction step, the error can accumulate over a long rollout and result in artificial heating, cooling, or loss of structural stability.

\subsubsection{Body Centered Cubic Iron}

\begin{figure}[H]
    \centering
    \includegraphics[width=1.0\linewidth]{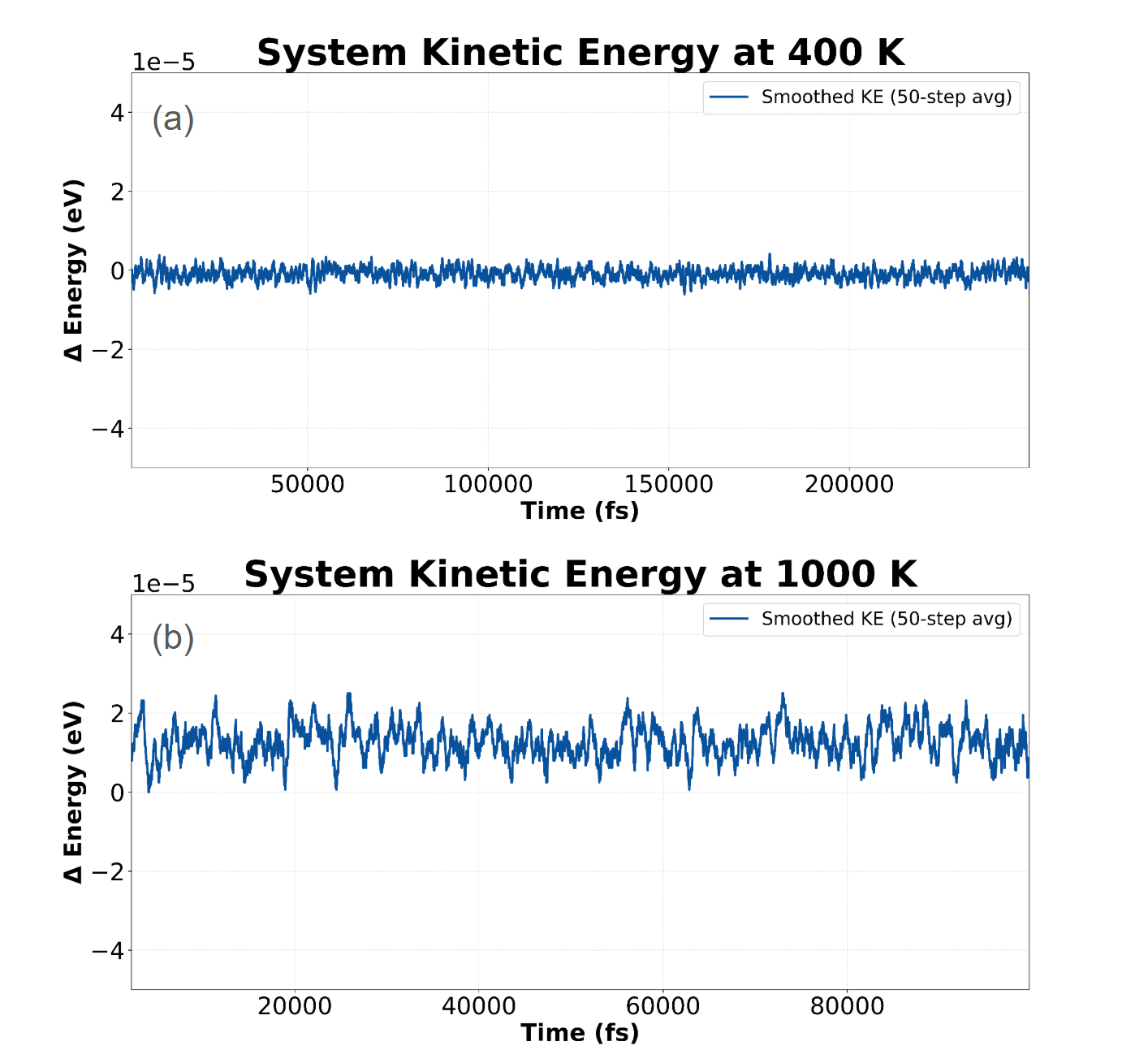}
    \caption{Deviation of the system kinetic energy from its reference value during autoregressive rollout of the iron model, smoothed with a 50-step moving average. (a) is a 250 ps rollout at 400~K. (b) is a 100 ps rollout at 1000~K.}
    \label{fig:fe-ke}
\end{figure}

\cref{fig:fe-ke} shows the kinetic energy deviation during autoregressive rollouts for BCC iron. At 400~K, the deviation remains centered near zero over 250~ps, with fluctuations of approximately $\pm 5 \times 10^{-6}$~eV. At 1000~K, the deviation fluctuates around approximately $-1.3 \times 10^{-5}$~eV over 100~ps. Although a small offset is present at the higher temperature, it remains bounded throughout the rollout and shows no systematic drift. The larger fluctuations at 1000~K are consistent with the increased thermal motion at higher temperature. The magnitude of the kinetic energy deviation remains on the order of $10^{-5}$~eV throughout these rollouts. More importantly, no progressive increase or decrease is observed with time, indicating that autoregressive error does not lead to systematic heating or cooling of the system. The absence of secular drift demonstrates that the surrogate maintains thermodynamic stability over extended propagation, avoiding the energy accumulation that can limit the long-time stability of learned atomistic simulators \cite{fu2023forces,stocker2022robust}.

\subsubsection{Hexagonal Close-Packed Magnesium}

\begin{figure}[H]
    \centering
    \includegraphics[width=1\linewidth]{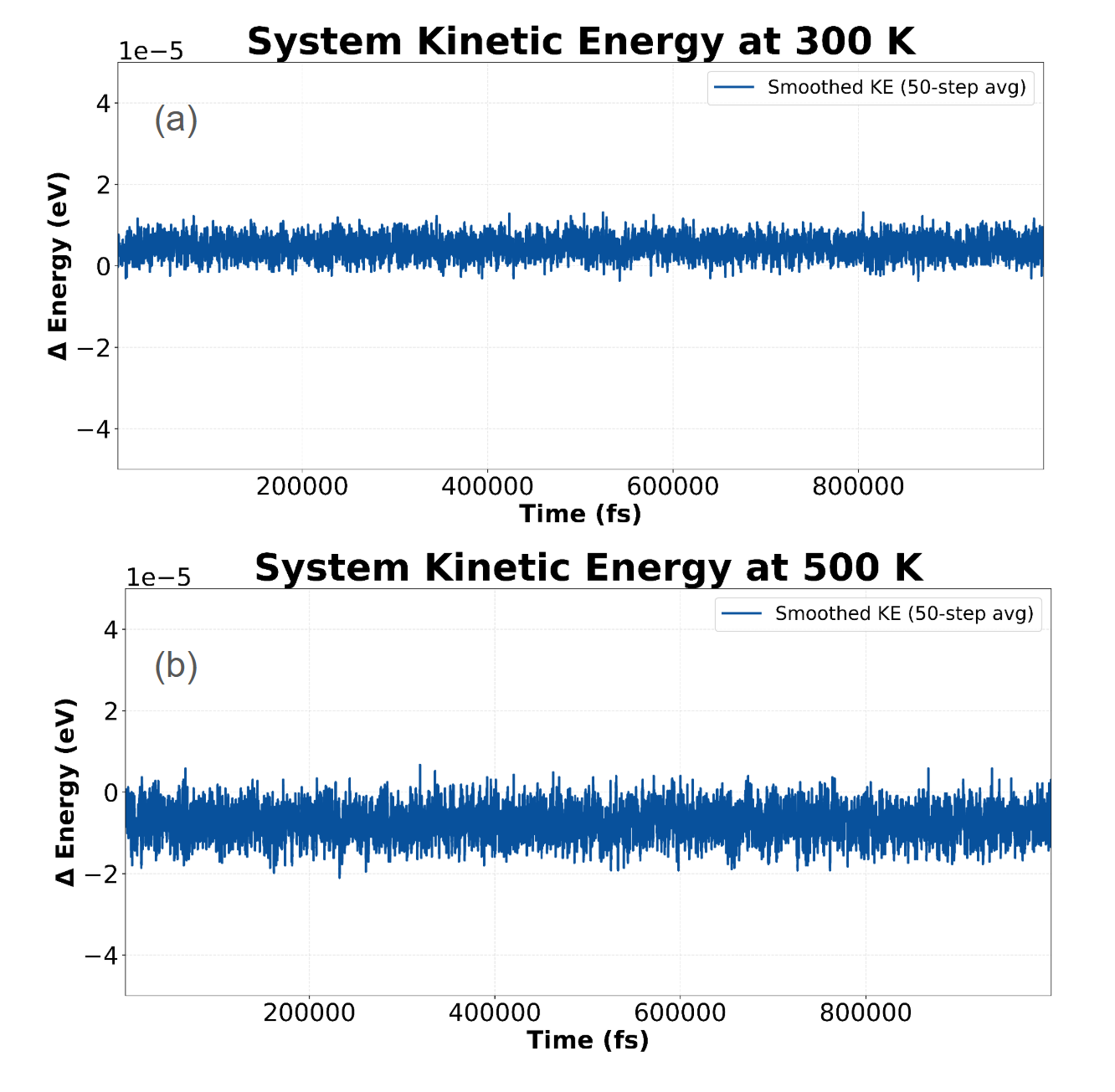}
    \caption{Deviation of the system kinetic energy from its reference value during 1 ns autoregressive rollouts of the magnesium model, smoothed with a 50-step moving average. (a) is at 300~K and (b) is at 500~K. Both rollouts correspond to approximately one million conventional integration steps, and neither exhibits secular drift.}
    \label{fig:mg-ke}
\end{figure}

\cref{fig:mg-ke} presents the kinetic energy deviation for magnesium over a 1~ns autoregressive rollout. At 300~K, Fig. 4(a), deviation fluctuates around approximately $5 \times 10^{-6}$~eV without a systematic increase or decrease over time. At 500~K, Fig. 4(b), the kinetic energy deviation fluctuates around approximately $-8 \times 10^{-6}$~eV. The opposite signs of the values around which the kinetic energy deviations fluctuate suggest that these differences are specific to the individual rollouts. The observed bias therefore cannot be attributed to a fixed tendency of the model to increase or decrease the kinetic energy. It may be noted that the one nanosecond trajectory corresponds to approximately one million integration steps in a conventional simulation with a one femtosecond time step. In the present work, the surrogate model predicts the atomic coordinates over a longer time interval at each step, which reduces the number of sequential prediction steps required to reach the same simulation time. In addition, no substantial difference is observed between the kinetic energy at the beginning and at the end of the rollout. Long-horizon stability in a non-cubic lattice was not examined in our earlier work on aluminum. The present results therefore provide additional evidence that stable autoregressive predictions can be maintained over extended simulation times.

\subsection{Structural Fidelity}
In addition to the kinetic energy analysis, the structural stability of the predicted trajectories was also examined. This examination is necessary because the lattice may distort or amorphize even when the kinetic energy remains stable over the rollout. The radial distribution function provides a direct measure for determining whether the correct crystal structure is maintained over the rollout.

\subsubsection{Body Centered Cubic Iron}

\begin{figure}[H]
    \centering
    \includegraphics[width=\linewidth]{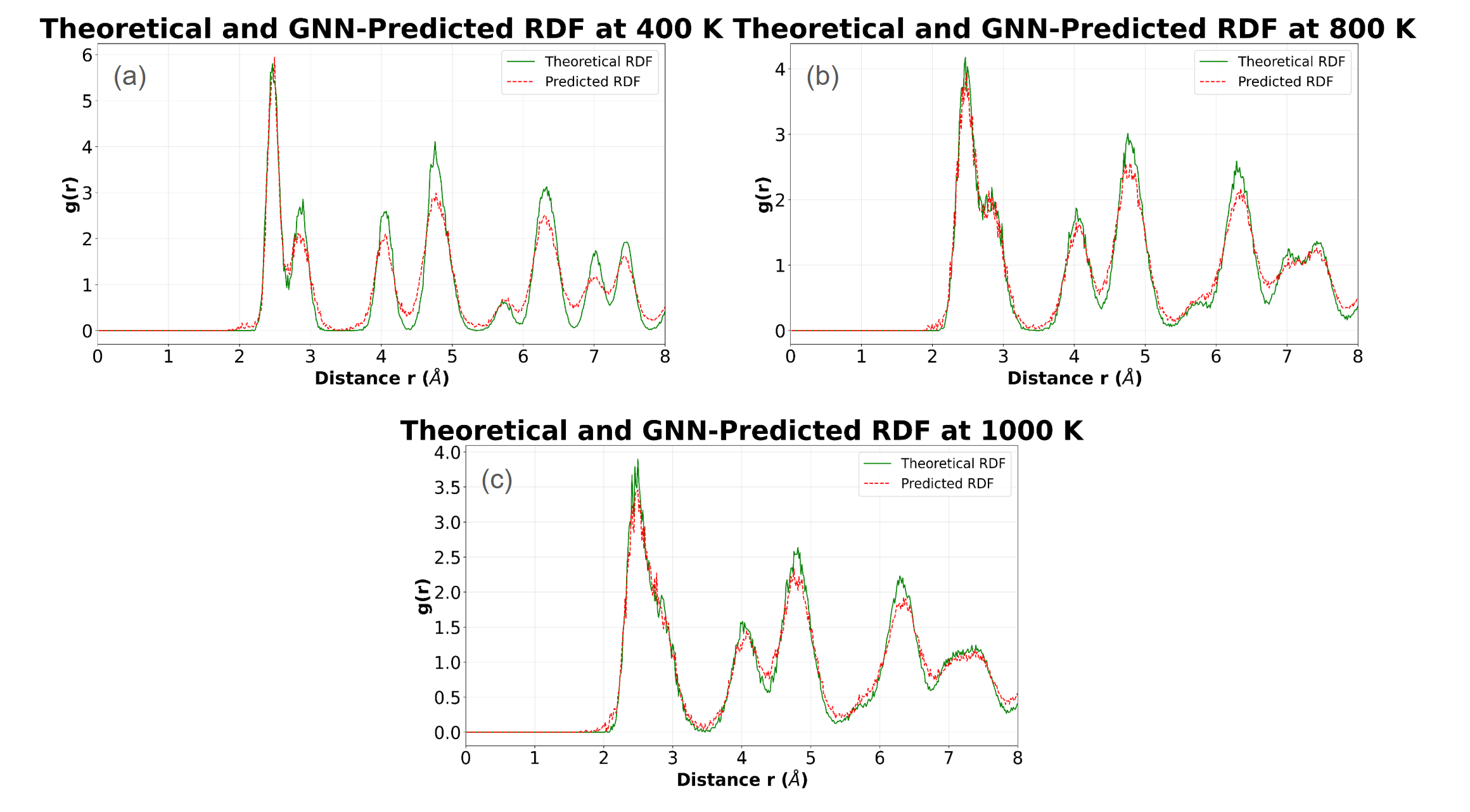}
    \caption{Radial distribution functions for iron from 20 ps rollouts.  (a) is at 400~K,  (b) at 800~K, and  (c) at 1000~K. Solid green curves are the reference molecular dynamics result and dashed red curves are the surrogate prediction. Peak positions are reproduced exactly at all three temperatures, and the agreement in peak height improves as temperature increases.}
    \label{fig:fe-rdf-20}
\end{figure}

\cref{fig:fe-rdf-20} the theoretical and predicted radial distribution functions for iron over a rollout of 20 ps. At 400~K, the first peak occurs at 2.48~\AA{} in both the reference and predicted RDFs, with a height of approximately 5.9. In both the reference and predicted RDFs, the second coordination shell appears as a shoulder near 2.87~\AA. Reproducing this shoulder is a meaningful test. The two shells are separated by less than $0.4$~\AA, well inside the width of a single radial basis function, so the model cannot distinguish them by distance encoding alone and must instead learn their relative populations of eight and six from the geometry of the local environment. The shoulder is reproduced in the predicted RDF, although its height is underestimated. Higher-order peaks at 4.05, 4.75, 6.3, 7.0, and 7.5~\AA{} appear at the correct positions in the prediction, with heights that are systematically low by roughly 20 to 30 percent at the largest separations. The most pronounced deviation is at the 4.75~\AA{} peak, where the predicted height of approximately 2.9 falls short of the reference value of 4.1. This underestimation is temperature dependent in an informative way. At 800~K, panel (b), the agreement improves markedly across the whole range, and at 1000~K, panel (c), the two curves are nearly indistinguishable out to 8~\AA. The trend is consistent with the sharpness of the features rather than with any failure of the physics. At 400~K the distribution consists of narrow, well separated peaks, and any small error in the predicted displacement distribution broadens them and lowers their maxima while conserving the area under each shell. At higher temperature the reference peaks are themselves broad, so the same absolute displacement error produces a much smaller relative distortion. The surrogate is therefore most accurate in exactly the regime where conventional molecular dynamics is most expensive to converge.

\begin{figure}[H]
    \centering
    \includegraphics[width=\linewidth]{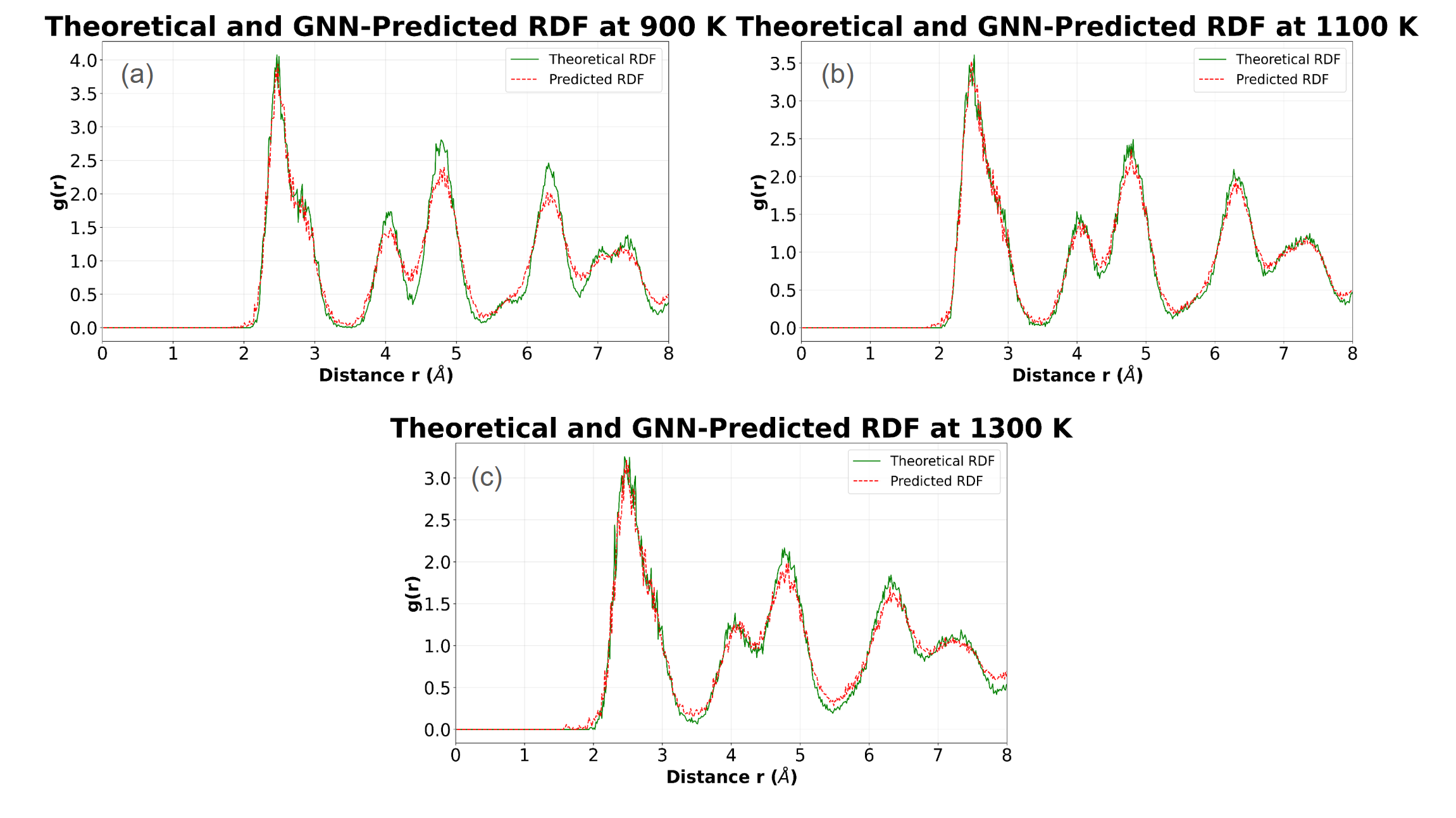}
    \caption{Radial distribution functions for iron from 100 ps rollouts.  (a) is at 900~K,  (b) at 1100~K, and  (c) at 1300~K. The body-centered cubic shell structure is preserved over the full rollout at all three temperatures, including at 1300~K.}
    \label{fig:fe-rdf-100}
\end{figure}

\cref{fig:fe-rdf-100} extends the rollout to 100 ps at 900, 1100, and 1300~K. Agreement is close at all three temperatures. At 900~K the predicted first peak reaches 3.95 against a reference value of 4.05, and at 1100~K the two first peaks coincide at 3.6. At 1300~K the predicted first peak of 3.1 sits slightly below the reference value of 3.25, and all higher shells overlap within the noise of the two curves. The 1300~K result deserves emphasis. The surrogate maintains the body-centered cubic coordination sequence at a temperature where the lattice is strongly anharmonic and thermal displacements are large. There is no evidence of the progressive loss of shell structure that would signal spurious amorphization, which is the characteristic failure mode of learned propagators pushed outside their training regime \cite{bihani2023stability}.

\subsubsection{Hexagonal Close-Packed Magnesium}

\begin{figure}[H]
    \centering
    \includegraphics[width=\linewidth]{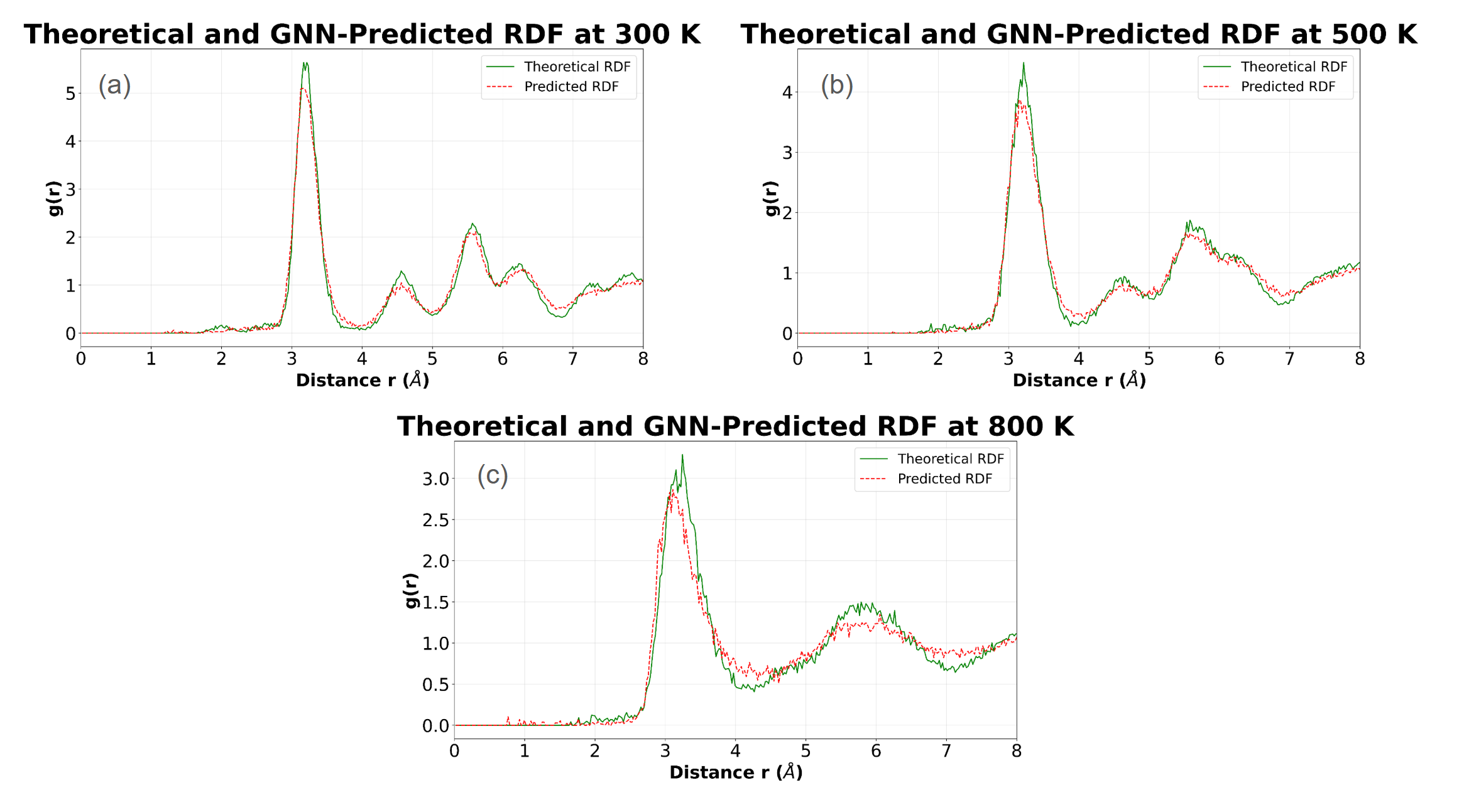}
    \caption{Radial distribution functions for magnesium from 20 ps rollouts.  (a) is at 300~K,  (b) at 500~K, and  (c) at 800~K. The hexagonal close-packed shell sequence is reproduced at all three temperatures.}
    \label{fig:mg-rdf-20}
\end{figure}

The magnesium results in \cref{fig:mg-rdf-20} show the same pattern in a different lattice. At 300~K, panel (a), the first peak sits at 3.2~\AA{} with a reference height of 5.6 and a predicted height of 5.1. The characteristic hexagonal close-packed shell sequence at 4.55, 5.55, and 6.3~\AA{} is present in both curves. The predicted curve slightly overestimates the minimum near 4.0~\AA{} and slightly underestimates the peak at 4.55~\AA, which together indicate a modest excess of intermediate-range disorder in the predicted configurations. At 500~K, panel (b), the first peak is predicted at 3.8 against a reference of 4.5, the largest relative deviation observed in the magnesium data. At 800~K, panel (c), agreement is again close, with the first peak predicted at 3.05 against a reference of 3.3 and with the broad second feature near 5.8~\AA{} well reproduced.

\begin{figure}[H]
    \centering
    \includegraphics[width=\linewidth]{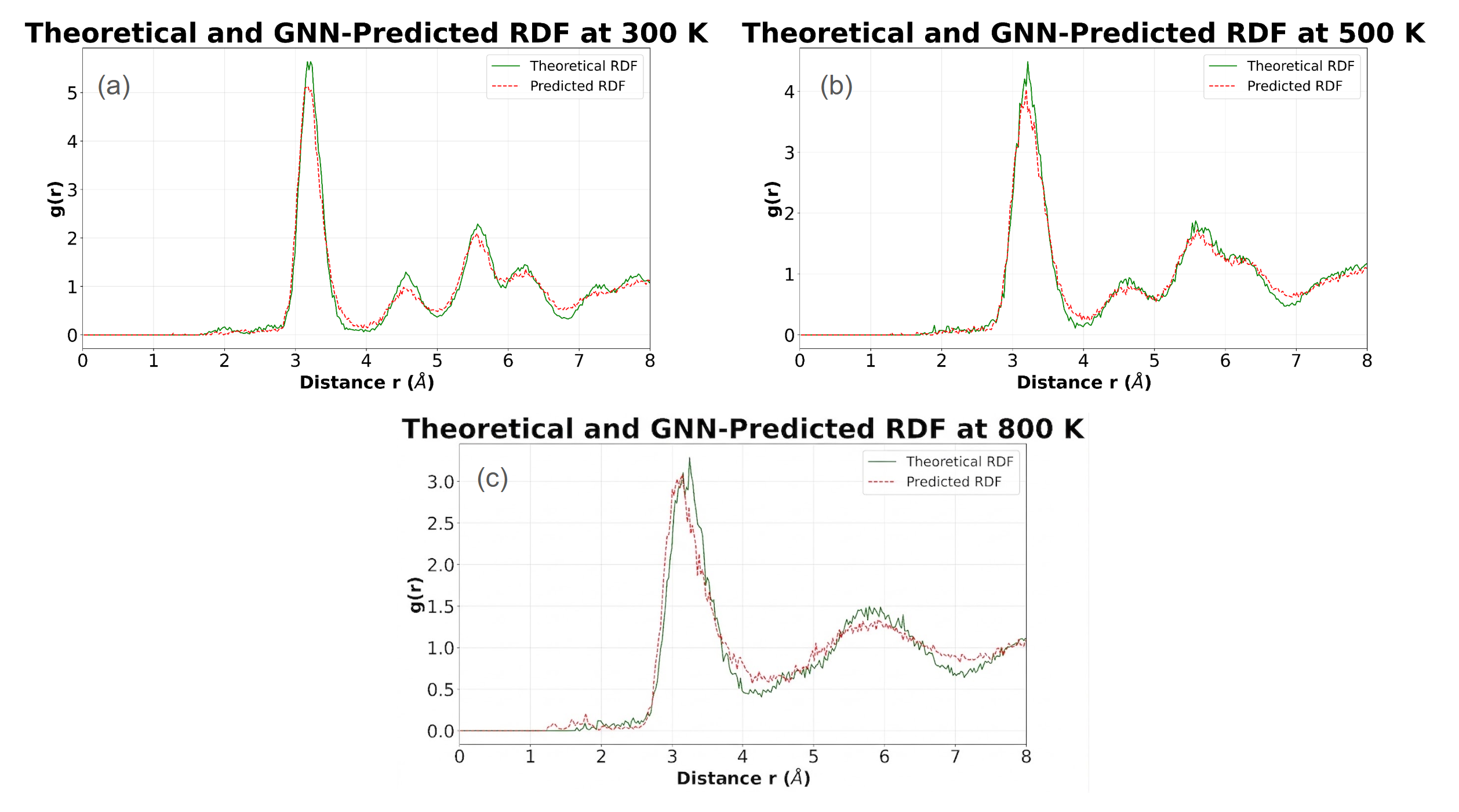}
    \caption{Radial distribution functions for magnesium from 100 ps rollouts at 300~K (a), 500~K (b), and 800~K (c). Structural agreement is maintained as the rollout is lengthened by a factor of five relative to \cref{fig:mg-rdf-20}.}
    \label{fig:mg-rdf-100}
\end{figure}

Lengthening the rollout to 100 ps, \cref{fig:mg-rdf-100}, produces distribution functions that are essentially unchanged from the 20 ps case at every temperature. The absence of any progressive degradation between the two figures is the point of the comparison. If the learned operator introduced a systematic bias in the displacement distribution, the error would compound over successive steps and the predicted shells would broaden monotonically with rollout length. They do not.

\begin{figure}[H]
    \centering
    \includegraphics[width=1\linewidth]{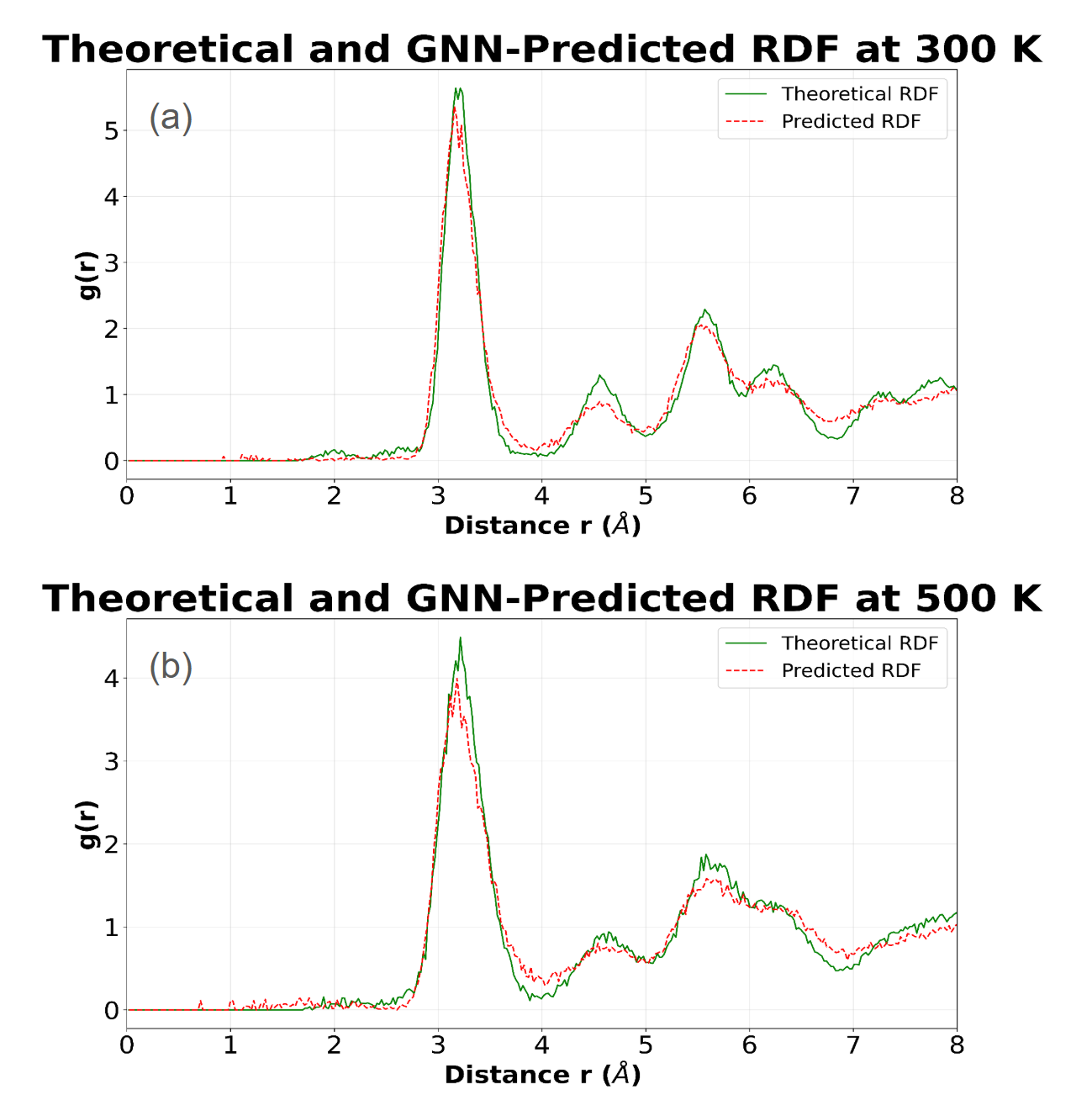}
    \caption{Radial distribution functions for magnesium from 1 ns rollouts at 300~K (a) and 500~K (b). The hexagonal close-packed structure is retained over approximately one million equivalent integration steps.}
    \label{fig:mg-rdf-1000}
\end{figure}

\cref{fig:mg-rdf-1000} carries the test to one nanosecond. At 300~K the predicted first peak reaches 5.35 against a reference value of 5.6, which is closer than the corresponding 20 ps result, and the shells at 4.55 and 5.55~\AA{} remain distinct. At 500~K the predicted first peak of 3.8 matches the 20 ps prediction, and the second shell structure is preserved. Taken with the kinetic energy traces in \cref{fig:mg-ke}, this establishes that the magnesium surrogate is stable in both the thermodynamic and the structural sense over a full nanosecond, which is the timescale at which surrogate models become practically useful.

\subsection{Dynamical Consistency}
The mean squared displacement provides an additional measure of dynamical consistency, as it is sensitive to the amplitude of atomic motion and not only to its time-averaged distribution.

\subsubsection{Iron}

\begin{figure}[H]
    \centering
    \includegraphics[width=\linewidth]{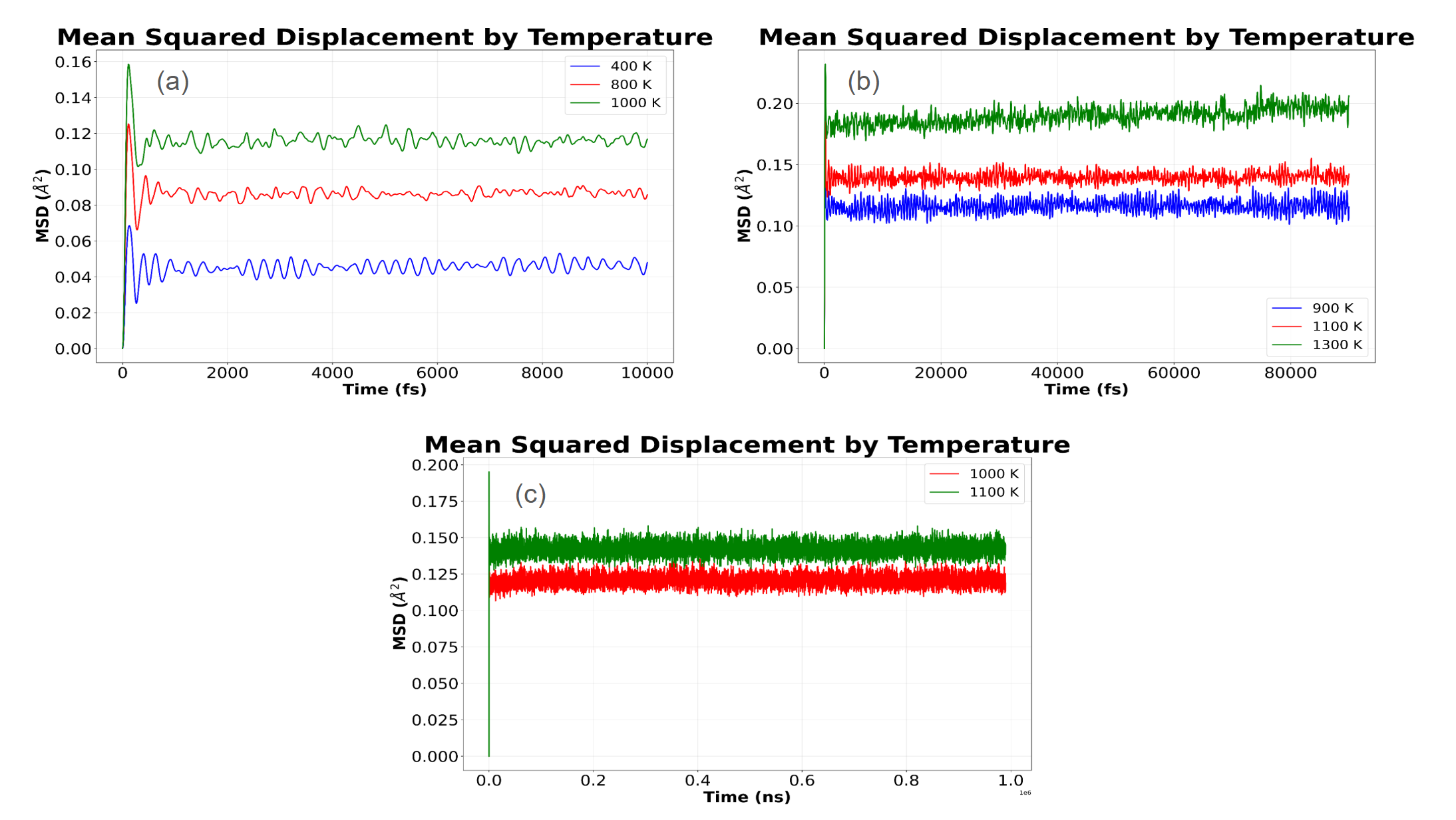}
    \caption{Mean squared displacement for iron.  (a) is a 10 ps rollout at 400, 800, and 1000~K.  (b) is a 100 ps rollout at 900, 1100, and 1300~K.  (c) is a 1 ns rollout at 1000 and 1100~K. All curves saturate at a plateau, and the plateau heights order correctly with temperature in all MSD calculation.}
    \label{fig:fe-msd}
\end{figure}

\cref{fig:fe-msd} shows mean squared displacements for iron. In panel (a), for the 10~ps rollouts, the mean squared displacement initially increases sharply and is followed by oscillations. The mean squared displacement then approaches approximately 0.045, 0.085, and 0.115~\AA$^2$ at 400, 800, and 1000~K, respectively. The oscillations are the coherent vibrational response of the lattice to the initialization of the trajectory, and their period corresponds to the characteristic phonon timescale. That the surrogate reproduces this transient is notable, since it is a collective dynamical feature that no per-atom displacement target contains explicitly.

The plateau heights order correctly and scale approximately linearly with temperature, as expected for a harmonic solid in which the mean squared displacement is proportional to $k_B T$ divided by an effective force constant. The ratio of the 1000~K plateau to the 400~K plateau is approximately 2.6, close to the ratio of temperatures of 2.5, and the slight excess is consistent with the anharmonic softening expected at elevated temperature. Panel (b) extends this to 900, 1100, and 1300~K over 100 ps, with plateaus at approximately 0.115, 0.14, and 0.19~\AA$^2$. Panel (c) reaches one nanosecond at 1000 and 1100~K, with plateaus at approximately 0.12 and 0.14~\AA$^2$. The 1000~K plateau of 0.12~\AA$^2$ in panel (c) agrees with the 0.115~\AA$^2$ obtained in panel (a) from a rollout one hundred times shorter, and the 1100~K plateau agrees between panels (b) and (c). This internal consistency across three independent rollouts of very different length is a strong indication that the learned dynamics are stationary.

It may be noted that no substantial increase in the mean squared displacement is observed after the plateau is reached. The absence of a sustained linear increase indicates that the surrogate does not produce spurious diffusion. For these defect-free crystals at temperatures below the melting point, the mean squared displacement is expected to approach a plateau. A small directional bias accumulated over the rollout would instead lead to a gradual increase in the mean squared displacement, similar to that associated with atomic diffusion. Such an increase is not observed, including at 1300~K.

\subsubsection{Magnesium}

\begin{figure}[H]
    \centering
    \includegraphics[width=1\linewidth]{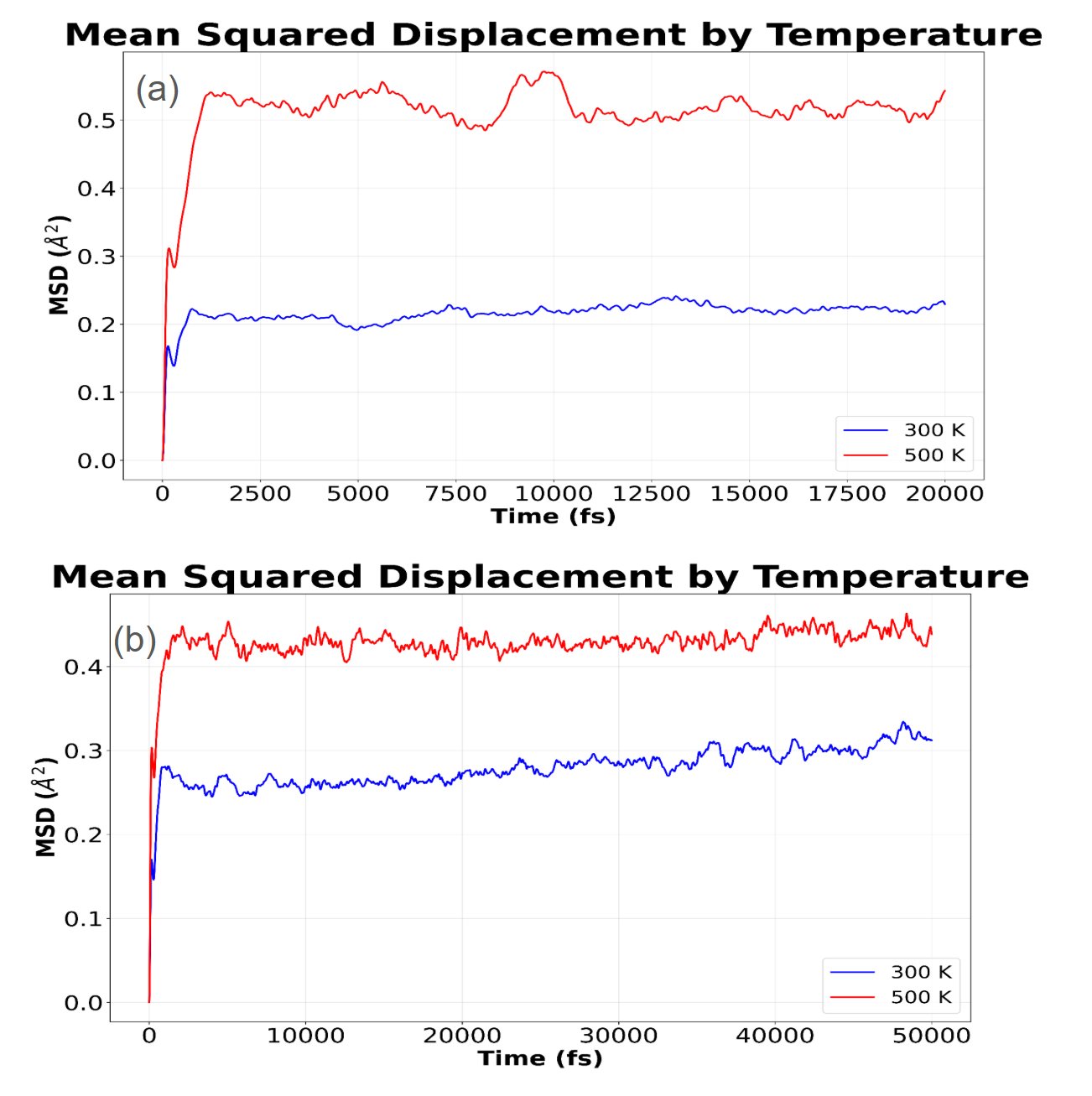}
    \caption{Mean squared displacement for magnesium.  (a) is a 20 ps rollout at 300 and 500~K.  (b) is a 50 ps rollout at 300 and 500~K. Plateau heights order correctly with temperature in both MSD calculations.}
    \label{fig:mg-msd}
\end{figure}

\cref{fig:mg-msd} shows the mean squared displacement for magnesium. Over 20~ps, in Fig. 11(a), the mean squared displacement approaches approximately 0.22~\AA$^2$ at 300~K and 0.51~\AA$^2$ at 500~K. Over 50~ps, Fig. 11(b), the mean squared displacement approaches approximately 0.27~\AA$^2$ at 300~K and 0.43~\AA$^2$ at 500~K. Two observations can be made from the mean squared displacement for magnesium. First, the magnesium plateaus are roughly four to five times larger than the iron plateaus at comparable temperature. This difference is consistent with the physical properties of the two materials. Magnesium has a much lower melting point and correspondingly smaller force constants, so at a given temperature the atoms undergo larger displacements around their lattice sites. The difference in absolute amplitude obtained with the same architecture trained separately for the two materials indicates that the surrogate captures the differences in their force constants. Second, for magnesium, a slight increase in the mean squared displacement is observed at 300~K in Fig. 11(b). This is a departure from ideal behavior. Over 50~ps, the mean squared displacement increases by approximately 15\% from its initial plateau. There are two possible explanations for this increase. The increase may reflect a small residual bias in the learned operator that accumulates slowly under autoregressive rollout. Alternatively it may reflect slow relaxation in the reference trajectory, since hexagonal close-packed lattices support low frequency basal modes that require long times to equilibrate. The one-nanosecond kinetic energy traces in \cref{fig:mg-ke} and the
one-nanosecond radial distribution functions in \cref{fig:mg-rdf-1000}
both remain stationary at 300~K. This suggests that the observed increase
may be associated with slow physical relaxation rather than systematic
error accumulation during the autoregressive rollout. However, determining
its origin would require direct comparison of the mean squared displacement
with a reference trajectory of matched length. The origin of this increase
therefore remains uncertain.

\subsection{Ablation Study}

To examine how different model components contribute to the predictions for the two lattice structures, we performed two ablation studies, in which selected components were removed and the models were retrained. For the feature ablation, the five per-atom structural descriptors were removed. Positions, velocities, atomic number, prediction horizon, and temperature were retained. The results are shown in \cref{tab:ablation}. 

\begin{table}[H]
\centering
\caption{Change in converged validation loss for each ablation relative to the full model, reported as a percentage increase. A larger increase indicates a more important component.}
\label{tab:ablation}
\begin{tabular}{lcc}
\toprule
\textbf{System} & \textbf{Feature ablation} & \textbf{Parameter ablation} \\
\midrule
Iron (body-centered cubic)        & $+9.7\%$ & $+37\%$ \\
Magnesium (hexagonal close-packed) & $+8.2\%$ & $+3.6\%$ \\
\bottomrule
\end{tabular}
\end{table}

The feature ablation results in comparable increases in prediction error for both systems: 9.7\% for iron and 8.2\% for magnesium. This indicates that the structural descriptors provide additional information. Removing the structural descriptors results in similar increases in prediction error for both crystal structures. The similar effect of the feature ablation for iron and magnesium indicates that the same set of structural descriptors can be used for both lattice structures.


For the model-capacity ablation, the hidden feature dimension was reduced from 128 to 64 while all other model and training parameters were held fixed. As shown in \cref{tab:ablation}, the effect of this reduction differs substantially between iron and magnesium. The validation loss increases by 37\% for iron, compared with only 3.6\% for magnesium. This difference may reflect the distinct coordination geometries of the two crystal structures. In body-centered cubic iron, the first and second neighbor shells occur at 2.48 and 2.87~\AA{}, respectively, and overlap substantially after thermal broadening. The two shells contain eight and six neighbors, respectively. Representing the contributions from these closely spaced coordination shells may require a higher model capacity. In hexagonal close-packed magnesium, the twelve nearest neighbors form a single shell at approximately 3.2~\AA{}, and the next shell occurs at approximately 4.55~\AA{}. The larger separation between the coordination shells may account for the smaller change in validation loss with the reduced model capacity. These results suggest that the sensitivity to model capacity may be related to differences in the coordination geometry of the two lattice structures.

\subsection{Transferability Across Crystal Symmetry}

The results above support a single conclusion that is broader than either metal individually. A direct position propagator, trained without symmetry-adapted bases and without any structure-specific architectural component, learns the evolution operator of face-centered cubic aluminum, body-centered cubic iron, and hexagonal close-packed magnesium using an identical pipeline. The two lattices differ in coordination number, in shell spacing, in the presence or absence of cubic symmetry, and in the stiffness of the underlying bonding. Convergence behavior, as seen in \cref{fig:fe-loss} and \cref{fig:mg-loss}, is nonetheless quantitatively similar, and the resulting rollouts are stable and structurally faithful in all cases.

We attribute this generality to two design choices. The first is that the graph is rebuilt at every frame from the current positions, so the model never assumes a fixed connectivity or a fixed coordination number. Whatever neighbors exist within the cutoff are the neighbors the model sees, and a change of lattice is simply a change in the statistics of that neighbor set. The second is that attention is restricted to graph edges but is otherwise unconstrained. The network is free to learn that a neighbor at 2.87~\AA{} in iron carries different information than a neighbor at 2.48~\AA, and separately to learn that a basal neighbor in magnesium carries different information than an out-of-plane neighbor, without either relation being imposed a priori. Symmetry is learned from data rather than built into the functional form. This is a different design philosophy from that of equivariant interatomic potentials \cite{batzner2022nequip, batatia2022mace, musaelian2023allegro}, in which symmetry is enforced architecturally, and the present results suggest that the learned approach is adequate for the specific task of short-horizon position propagation in crystalline metals.

\section{Conclusions}

This work demonstrates the transferability of a transformer graph neural network surrogate for direct atomic position propagation across distinct crystal structures and elemental systems. Using the same model formulation, feature representation, graph construction, and training strategy, the framework was applied to BCC iron and HCP magnesium without introducing symmetry-specific modifications. The models achieved mean squared position errors on the order of $10^{-4}$~\AA$^2$ and remained stable during autoregressive rollouts extending to 1~ns, corresponding to approximately one million conventional molecular dynamics integration steps. Kinetic-energy deviations remained on the order of $10^{-5}$~eV without systematic secular drift.

The predicted trajectories also retained the characteristic structural and dynamical signatures of both lattices. For BCC iron, the surrogate reproduced the closely spaced first and second coordination shells at 2.48 and 2.87~\AA{}, respectively, and preserved the BCC coordination structure up to 1300~K. The HCP coordination structure of magnesium was similarly maintained during rollouts reaching 1~ns. Mean squared displacement curves exhibited temperature-dependent plateaus with distinct magnitudes for iron and magnesium, reflecting differences in their lattice dynamics. Ablation of the structural descriptors increased the validation loss by 9.7\% for iron and 8.2\% for magnesium, while reducing the hidden feature dimension from 128 to 64 increased the loss by 37\% and 3.6\%, respectively, indicating different sensitivities to model capacity across the two coordination environments. Together with its applicability to FCC aluminum, these results demonstrate that the direct-position GNN formulation is not restricted to a particular crystal symmetry and provides a common framework for learning atomic evolution across FCC, BCC, and HCP systems. The present study establishes transferability at the framework level, with separate models trained for each material. Extending the approach to jointly trained multielement systems, defects, interfaces, and phase transformations represents an important next step toward more general surrogate models for accelerated atomistic simulation.

\section*{Code Availability}

Workflows developed in this study are openly available at \url{https://github.com/mahata-lab/GNN-Molecular-Dynamics}.

\section*{Acknowledgments}

The authors acknowledge the use of computational resources at the Massachusetts Green High Performance Computing Center. This research also benefited from high performance computing allocations provided by the National Science Foundation through ACCESS under awards MAT250103 and MAT240094. Additional computational resources were supported by Argonne National Laboratory under the Director's Discretionary allocation for the project \textit{GNNMD}. Further support was provided through a National Science Foundation Major Research Instrumentation award to Wilkes University under award number 1920129, which contributed essential computational infrastructure for this study.

\bibliographystyle{unsrt}
\bibliography{references}

@article{bartok2013soap,
  author  = {Bart{\'o}k, Albert P. and Kondor, Risi and Cs{\'a}nyi, G{\'a}bor},
  title   = {On Representing Chemical Environments},
  journal = {Physical Review B},
  volume  = {87},
  number  = {18},
  pages   = {184115},
  year    = {2013},
  doi      = {10.1103/PhysRevB.87.184115}
}

@inproceedings{sanchez2020learning,
  author    = {Sanchez-Gonzalez, Alvaro and Godwin, Jonathan and Pfaff, Tobias and Ying, Rex and Leskovec, Jure and Battaglia, Peter W.},
  title     = {Learning to Simulate Complex Physics with Graph Networks},
  booktitle = {Proceedings of the 37th International Conference on Machine Learning (ICML)},
  pages     = {8459--8468},
  year      = {2020}
}

@book{allen_tildesley,
  author    = {Allen, Michael P. and Tildesley, Dominic J.},
  title     = {Computer Simulation of Liquids},
  edition   = {2},
  publisher = {Oxford University Press},
  address   = {Oxford, UK},
  year      = {2017}
}

@book{frenkel_smit,
  author    = {Frenkel, Daan and Smit, Berend},
  title     = {Understanding Molecular Simulation: From Algorithms to Applications},
  edition   = {2},
  publisher = {Academic Press},
  address   = {San Diego, CA},
  year      = {2002}
}

@book{rapaport,
  author    = {Rapaport, Dennis C.},
  title     = {The Art of Molecular Dynamics Simulation},
  edition   = {2},
  publisher = {Cambridge University Press},
  address   = {Cambridge, UK},
  year      = {2004}
}

@article{bitzek_deformation,
  author  = {Bitzek, Erik and Koskinen, Pekka and G{\"a}hler, Franz and Moseler, Michael and Gumbsch, Peter},
  title   = {Structural Relaxation Made Simple},
  journal = {Physical Review Letters},
  volume  = {97},
  number  = {17},
  pages   = {170201},
  year    = {2006},
  doi     = {10.1103/PhysRevLett.97.170201}
}

@book{marx_hutter,
  author    = {Marx, Dominik and Hutter, J{\"u}rg},
  title     = {Ab Initio Molecular Dynamics: Basic Theory and Advanced Methods},
  publisher = {Cambridge University Press},
  address   = {Cambridge, UK},
  year      = {2009}
}

@article{kresse_furthmuller,
  author  = {Kresse, G. and Furthm{\"u}ller, J.},
  title   = {Efficient Iterative Schemes for Ab Initio Total-Energy Calculations Using a Plane-Wave Basis Set},
  journal = {Physical Review B},
  volume  = {54},
  number  = {16},
  pages   = {11169--11186},
  year    = {1996},
  doi      = {10.1103/PhysRevB.54.11169}
}

@inproceedings{gilmer2017mpnn,
  author    = {Gilmer, Justin and Schoenholz, Samuel S. and Riley, Patrick F. and Vinyals, Oriol and Dahl, George E.},
  title     = {Neural Message Passing for Quantum Chemistry},
  booktitle = {Proceedings of the 34th International Conference on Machine Learning (ICML)},
  pages     = {1263--1272},
  year      = {2017}
}

@inproceedings{schuett2017schnet,
  author    = {Sch{\"u}tt, Kristof T. and Kindermans, Pieter-Jan and Sauceda, Huziel E. and Chmiela, Stefan and Tkatchenko, Alexandre and M{\"u}ller, Klaus-Robert},
  title     = {{SchNet}: A Continuous-Filter Convolutional Neural Network for Modeling Quantum Interactions},
  booktitle = {Advances in Neural Information Processing Systems (NeurIPS)},
  volume    = {30},
  year      = {2017}
}

@article{wang2018deepmd,
  author  = {Wang, Han and Zhang, Linfeng and Han, Jiequn and E, Weinan},
  title   = {{DeePMD-kit}: A Deep Learning Package for Many-Body Potential Energy Representation and Molecular Dynamics},
  journal = {Computer Physics Communications},
  volume  = {228},
  pages   = {178--184},
  year    = {2018}
}

@inproceedings{batatia2022mace,
  author    = {Batatia, Ilyes and Kov{\'a}cs, D{\'a}vid P{\'e}ter and Simm, Gregor N. C. and Ortner, Christoph and Cs{\'a}nyi, G{\'a}bor},
  title     = {{MACE}: Higher Order Equivariant Message Passing Neural Networks for Fast and Accurate Force Fields},
  booktitle = {Advances in Neural Information Processing Systems (NeurIPS)},
  volume    = {35},
  year      = {2022}
}

@article{mishin1999interatomic,
  author  = {Mishin, Y. and Farkas, D. and Mehl, M. J. and Papaconstantopoulos, D. A.},
  title   = {Interatomic Potentials for Monoatomic Metals from Experimental Data and Ab Initio Calculations},
  journal = {Physical Review B},
  volume  = {59},
  number  = {5},
  pages   = {3393--3407},
  year    = {1999},
  doi      = {10.1103/PhysRevB.59.3393}
}

@article{battaglia2018relational,
  author  = {Battaglia, Peter W. and Hamrick, Jessica B. and Bapst, Victor and Sanchez-Gonzalez, Alvaro and Zambaldi, Vinicius and Malinowski, Mateusz and Tacchetti, Andrea and Raposo, David and Santoro, Adam and Faulkner, Ryan and others},
  title   = {Relational Inductive Biases, Deep Learning, and Graph Networks},
  journal = {arXiv preprint arXiv:1806.01261},
  year    = {2018}
}

@inproceedings{velickovic2018gat,
  author    = {Veli{\v{c}}kovi{\'c}, Petar and Cucurull, Guillem and Casanova, Arantxa and Romero, Adriana and Li{\`o}, Pietro and Bengio, Yoshua},
  title     = {Graph Attention Networks},
  booktitle = {International Conference on Learning Representations (ICLR)},
  year      = {2018}
}

@article{thompson2022lammps,
  author  = {Thompson, Aidan P. and Aktulga, H. Metin and Berger, Richard and Bolintineanu, Dan S. and Brown, W. Michael and Crozier, Paul S. and in 't Veld, Pieter J. and Kohlmeyer, Axel and Moore, Stan G. and Nguyen, Trung Dac and Shan, Ray and Stevens, Mark J. and Tranchida, Julien and Trott, Christian and Plimpton, Steven J.},
  title   = {{LAMMPS}: A Flexible Simulation Tool for Particle-Based Materials Modeling at the Atomic, Meso, and Continuum Scales},
  journal = {Computer Physics Communications},
  volume  = {271},
  pages   = {108171},
  year    = {2022},
  doi      = {10.1016/j.cpc.2021.108171}
}

@article{kadau2006molecular,
  title={Molecular dynamics comes of age: 320 billion atom simulation on BlueGene/L},
  author={Kadau, Kai and Germann, Timothy C and Lomdahl, Peter S},
  journal={International Journal of Modern Physics C},
  volume={17},
  number={12},
  pages={1755--1761},
  year={2006},
  publisher={World Scientific}
}

@article{behler2007generalized,
  author  = {Behler, J{\"o}rg and Parrinello, Michele},
  title   = {Generalized Neural-Network Representation of High-Dimensional Potential-Energy Surfaces},
  journal = {Physical Review Letters},
  volume  = {98},
  number  = {14},
  pages   = {146401},
  year    = {2007},
  doi      = {10.1103/PhysRevLett.98.146401}
}

@article{batzner2022nequip,
  author  = {Batzner, Simon and Musaelian, Albert and Sun, Lixin and Geiger, Mario and Mailoa, Jonathan P. and Kornbluth, Mordechai and Molinari, Nicola and Smidt, Tess E. and Kozinsky, Boris},
  title   = {{E(3)}-Equivariant Graph Neural Networks for Data-Efficient and Accurate Interatomic Potentials},
  journal = {Nature Communications},
  volume  = {13},
  number  = {1},
  pages   = {2453},
  year    = {2022},
  doi      = {10.1038/s41467-022-29939-5}
}

@article{verlet1967,
  author  = {Verlet, Loup},
  title   = {Computer Experiments on Classical Fluids. I. Thermodynamical Properties of Lennard-Jones Molecules},
  journal = {Physical Review},
  volume  = {159},
  number  = {1},
  pages   = {98--103},
  year    = {1967}
}

@article{swope1982,
  author  = {Swope, William C. and Andersen, Hans C. and Berens, Peter H. and Wilson, Kent R.},
  title   = {A Computer Simulation Method for the Calculation of Equilibrium Constants for the Formation of Physical Clusters of Molecules},
  journal = {The Journal of Chemical Physics},
  volume  = {76},
  number  = {1},
  pages   = {637--649},
  year    = {1982}
}

@article{nose1984,
  author  = {Nos{\'e}, Shuichi},
  title   = {A Unified Formulation of the Constant Temperature Molecular Dynamics Methods},
  journal = {The Journal of Chemical Physics},
  volume  = {81},
  number  = {1},
  pages   = {511--519},
  year    = {1984}
}

@article{hoover1985,
  author  = {Hoover, William G.},
  title   = {Canonical Dynamics: Equilibrium Phase-Space Distributions},
  journal = {Physical Review A},
  volume  = {31},
  number  = {3},
  pages   = {1695--1697},
  year    = {1985}
}

@article{plimpton1995,
  author  = {Plimpton, Steve},
  title   = {Fast Parallel Algorithms for Short-Range Molecular Dynamics},
  journal = {Journal of Computational Physics},
  volume  = {117},
  number  = {1},
  pages   = {1--19},
  year    = {1995}
}

@article{stukowski2010ovito,
  author  = {Stukowski, Alexander},
  title   = {Visualization and Analysis of Atomistic Simulation Data with {OVITO}, the Open Visualization Tool},
  journal = {Modelling and Simulation in Materials Science and Engineering},
  volume  = {18},
  number  = {1},
  pages   = {015012},
  year    = {2010}
}

@article{daw1984eam,
  author  = {Daw, Murray S. and Baskes, M. I.},
  title   = {Embedded-Atom Method: Derivation and Application to Impurities, Surfaces, and Other Defects in Metals},
  journal = {Physical Review B},
  volume  = {29},
  number  = {12},
  pages   = {6443--6453},
  year    = {1984}
}

@article{finnis1984,
  author  = {Finnis, M. W. and Sinclair, J. E.},
  title   = {A Simple Empirical N-Body Potential for Transition Metals},
  journal = {Philosophical Magazine A},
  volume  = {50},
  number  = {1},
  pages   = {45--55},
  year    = {1984}
}

@article{hohenberg1964,
  author  = {Hohenberg, P. and Kohn, W.},
  title   = {Inhomogeneous Electron Gas},
  journal = {Physical Review},
  volume  = {136},
  number  = {3B},
  pages   = {B864--B871},
  year    = {1964}
}

@article{kohn1965,
  author  = {Kohn, W. and Sham, L. J.},
  title   = {Self-Consistent Equations Including Exchange and Correlation Effects},
  journal = {Physical Review},
  volume  = {140},
  number  = {4A},
  pages   = {A1133--A1138},
  year    = {1965}
}

@inproceedings{jia2020pushing,
  author    = {Jia, Weile and Wang, Han and Chen, Mohan and Lu, Denghui and Lin, Lin and Car, Roberto and E, Weinan and Zhang, Linfeng},
  title     = {Pushing the Limit of Molecular Dynamics with Ab Initio Accuracy to 100 Million Atoms with Machine Learning},
  booktitle = {Proceedings of the International Conference for High Performance Computing, Networking, Storage and Analysis (SC20)},
  pages     = {1--14},
  year      = {2020}
}

@article{voter2002extending,
  author  = {Voter, Arthur F. and Montalenti, Francesco and Germann, Timothy C.},
  title   = {Extending the Time Scale in Atomistic Simulation of Materials},
  journal = {Annual Review of Materials Research},
  volume  = {32},
  pages   = {321--346},
  year    = {2002}
}

@article{perez2009accelerated,
  author  = {Perez, Danny and Uberuaga, Blas P. and Shim, Yunsic and Amar, Jacques G. and Voter, Arthur F.},
  title   = {Accelerated Molecular Dynamics Methods: Introduction and Recent Developments},
  journal = {Annual Reports in Computational Chemistry},
  volume  = {5},
  pages   = {79--98},
  year    = {2009}
}

@article{honeycutt1987cna,
  author  = {Honeycutt, J. Dana and Andersen, Hans C.},
  title   = {Molecular Dynamics Study of Melting and Freezing of Small Lennard-Jones Clusters},
  journal = {The Journal of Physical Chemistry},
  volume  = {91},
  number  = {19},
  pages   = {4950--4963},
  year    = {1987}
}

@article{kelchner1998csp,
  author  = {Kelchner, Cynthia L. and Plimpton, S. J. and Hamilton, J. C.},
  title   = {Dislocation Nucleation and Defect Structure During Surface Indentation},
  journal = {Physical Review B},
  volume  = {58},
  number  = {17},
  pages   = {11085--11088},
  year    = {1998}
}

@article{steinhardt1983bond,
  author  = {Steinhardt, Paul J. and Nelson, David R. and Ronchetti, Marco},
  title   = {Bond-Orientational Order in Liquids and Glasses},
  journal = {Physical Review B},
  volume  = {28},
  number  = {2},
  pages   = {784--805},
  year    = {1983}
}

@article{mickel2013minkowski,
  author  = {Mickel, Walter and Kapfer, Sebastian C. and Schr{\"o}der-Turk, Gerd E. and Mecke, Klaus},
  title   = {Shortcomings of the Bond Orientational Order Parameters for the Analysis of Disordered Particulate Matter},
  journal = {The Journal of Chemical Physics},
  volume  = {138},
  number  = {4},
  pages   = {044501},
  year    = {2013}
}

@article{lechner2008accurate,
  author  = {Lechner, Wolfgang and Dellago, Christoph},
  title   = {Accurate Determination of Crystal Structures Based on Averaged Local Bond Order Parameters},
  journal = {The Journal of Chemical Physics},
  volume  = {129},
  number  = {11},
  pages   = {114707},
  year    = {2008}
}

@inproceedings{vaswani2017attention,
  author    = {Vaswani, Ashish and Shazeer, Noam and Parmar, Niki and Uszkoreit, Jakob and Jones, Llion and Gomez, Aidan N. and Kaiser, {\L}ukasz and Polosukhin, Illia},
  title     = {Attention Is All You Need},
  booktitle = {Advances in Neural Information Processing Systems (NeurIPS)},
  volume    = {30},
  year      = {2017}
}

@inproceedings{shi2021masked,
  author    = {Shi, Yunsheng and Huang, Zhengjie and Feng, Shikun and Zhong, Hui and Wang, Wenjin and Sun, Yu},
  title     = {Masked Label Prediction: Unified Message Passing Model for Semi-Supervised Classification},
  booktitle = {Proceedings of the Thirtieth International Joint Conference on Artificial Intelligence (IJCAI)},
  pages     = {1548--1554},
  year      = {2021}
}

@article{musaelian2023allegro,
  author  = {Musaelian, Albert and Batzner, Simon and Johansson, Anders and Sun, Lixin and Owen, Cameron J. and Kornbluth, Mordechai and Kozinsky, Boris},
  title   = {Learning Local Equivariant Representations for Large-Scale Atomistic Dynamics},
  journal = {Nature Communications},
  volume  = {14},
  number  = {1},
  pages   = {579},
  year    = {2023}
}

@article{chen2022m3gnet,
  author  = {Chen, Chi and Ong, Shyue Ping},
  title   = {A Universal Graph Deep Learning Interatomic Potential for the Periodic Table},
  journal = {Nature Computational Science},
  volume  = {2},
  number  = {11},
  pages   = {718--728},
  year    = {2022}
}

@article{deng2023chgnet,
  author  = {Deng, Bowen and Zhong, Peichen and Jun, KyuJung and Riebesell, Janosh and Han, Kevin and Bartel, Christopher J. and Ceder, Gerbrand},
  title   = {{CHGNet} as a Pretrained Universal Neural Network Potential for Charge-Informed Atomistic Modelling},
  journal = {Nature Machine Intelligence},
  volume  = {5},
  number  = {9},
  pages   = {1031--1041},
  year    = {2023}
}

@article{merchant2023gnome,
  author  = {Merchant, Amil and Batzner, Simon and Schoenholz, Samuel S. and Aykol, Muratahan and Cheon, Gowoon and Cubuk, Ekin Dogus},
  title   = {Scaling Deep Learning for Materials Discovery},
  journal = {Nature},
  volume  = {624},
  number  = {7990},
  pages   = {80--85},
  year    = {2023}
}

@article{unke2021review,
  author  = {Unke, Oliver T. and Chmiela, Stefan and Sauceda, Huziel E. and Gastegger, Michael and Poltavsky, Igor and Sch{\"u}tt, Kristof T. and Tkatchenko, Alexandre and M{\"u}ller, Klaus-Robert},
  title   = {Machine Learning Force Fields},
  journal = {Chemical Reviews},
  volume  = {121},
  number  = {16},
  pages   = {10142--10186},
  year    = {2021}
}

@article{zuo2020performance,
  author  = {Zuo, Yunxing and Chen, Chi and Li, Xiangguo and Deng, Zhi and Chen, Yiming and Behler, J{\"o}rg and Cs{\'a}nyi, G{\'a}bor and Shapeev, Alexander V. and Thompson, Aidan P. and Wood, Mitchell A. and Ong, Shyue Ping},
  title   = {Performance and Cost Assessment of Machine Learning Interatomic Potentials},
  journal = {The Journal of Physical Chemistry A},
  volume  = {124},
  number  = {4},
  pages   = {731--745},
  year    = {2020}
}

@article{batatia2023foundation,
  author  = {Batatia, Ilyes and Benner, Philipp and Chiang, Yuan and Elena, Alin M. and Kov{\'a}cs, D{\'a}vid P. and Riebesell, Janosh and Advincula, Xavier R. and Asta, Mark and Baldwin, William J. and Bernstein, Noam and others},
  title   = {A Foundation Model for Atomistic Materials Chemistry},
  journal = {arXiv preprint arXiv:2401.00096},
  year    = {2023}
}

@article{fu2023forces,
  author  = {Fu, Xiang and Wu, Zhenghao and Wang, Wujie and Xie, Tian and Keten, Sinan and Gomez-Bombarelli, Rafael and Jaakkola, Tommi},
  title   = {Forces Are Not Enough: Benchmark and Critical Evaluation for Machine Learning Force Fields with Molecular Simulations},
  journal = {Transactions on Machine Learning Research},
  year    = {2023}
}

@article{stocker2022robust,
  author  = {Stocker, Sina and Gasteiger, Johannes and Becker, Florian and G{\"u}nnemann, Stephan and Margraf, Johannes T.},
  title   = {How Robust Are Modern Graph Neural Network Potentials in Long and Hot Molecular Dynamics Simulations?},
  journal = {Machine Learning: Science and Technology},
  volume  = {3},
  number  = {4},
  pages   = {045010},
  year    = {2022}
}

@article{bihani2023stability,
  author  = {Bihani, Vaibhav and Mannan, Sajid and Pratiush, Utkarsh and Du, Tao and Chen, Zhimin and Miret, Santiago and Micoulaut, Matthieu and Smedskjaer, Morten M. and Ranu, Sayan and Krishnan, N. M. Anoop},
  title   = {{EGraFFBench}: Evaluation of Equivariant Graph Neural Network Force Fields for Atomistic Simulations},
  journal = {Digital Discovery},
  volume  = {3},
  pages   = {759--768},
  year    = {2024}
}

@inproceedings{pfaff2021meshgraphnets,
  author    = {Pfaff, Tobias and Fortunato, Meire and Sanchez-Gonzalez, Alvaro and Battaglia, Peter W.},
  title     = {Learning Mesh-Based Simulation with Graph Networks},
  booktitle = {International Conference on Learning Representations (ICLR)},
  year      = {2021}
}

@article{klein2023timewarp,
  author  = {Klein, Leon and Foong, Andrew Y. K. and Fjelde, Tor Erlend and Mlodozeniec, Bruno and Brockschmidt, Marc and Nowozin, Sebastian and No{\'e}, Frank and Tomioka, Ryota},
  title   = {Timewarp: Transferable Acceleration of Molecular Dynamics by Learning Time-Coarsened Dynamics},
  journal = {Advances in Neural Information Processing Systems (NeurIPS)},
  volume  = {36},
  year    = {2023}
}

@article{hsu2024score,
  author  = {Hsu, Tim and Sadigh, Babak and Bulatov, Vasily and Zhou, Fei},
  title   = {Score Dynamics: Scaling Molecular Dynamics with Picoseconds Time Steps via Conditional Diffusion Model},
  journal = {Journal of Chemical Theory and Computation},
  volume  = {20},
  number  = {6},
  pages   = {2335--2348},
  year    = {2024}
}

@article{jing2024mdgen,
  author  = {Jing, Bowen and Stark, Hannes and Jaakkola, Tommi and Berger, Bonnie},
  title   = {Generative Modeling of Molecular Dynamics Trajectories},
  journal = {Advances in Neural Information Processing Systems (NeurIPS)},
  volume  = {37},
  year    = {2024}
}

@article{vlachas2022multiscale,
  author  = {Vlachas, Pantelis R. and Zavadlav, Julija and Praprotnik, Matej and Koumoutsakos, Petros},
  title   = {Accelerated Simulations of Molecular Systems Through Learning of Effective Dynamics},
  journal = {Journal of Chemical Theory and Computation},
  volume  = {18},
  number  = {1},
  pages   = {538--549},
  year    = {2022}
}

@article{marian2004dynamic,
  author  = {Marian, Jaime and Cai, Wei and Bulatov, Vasily V.},
  title   = {Dynamic Transitions from Smooth to Rough to Twinning in Dislocation Motion},
  journal = {Nature Materials},
  volume  = {3},
  number  = {3},
  pages   = {158--163},
  year    = {2004}
}

@article{gilbert2008structure,
  author  = {Gilbert, M. R. and Queyreau, S. and Marian, J.},
  title   = {Stress and Temperature Dependence of Screw Dislocation Mobility in $\alpha$-Fe by Molecular Dynamics},
  journal = {Physical Review B},
  volume  = {84},
  number  = {17},
  pages   = {174103},
  year    = {2011}
}

@article{dudarev2013density,
  author  = {Dudarev, S. L.},
  title   = {Density Functional Theory Models for Radiation Damage},
  journal = {Annual Review of Materials Research},
  volume  = {43},
  pages   = {35--61},
  year    = {2013}
}

@article{malerba2010comparison,
  author  = {Malerba, L. and Marinica, M. C. and Anento, N. and Bj{\"o}rkas, C. and Nguyen, H. and Domain, C. and Djurabekova, F. and Olsson, P. and Nordlund, K. and Serra, A. and Terentyev, D. and Willaime, F. and Becquart, C. S.},
  title   = {Comparison of Empirical Interatomic Potentials for Iron Applied to Radiation Damage Studies},
  journal = {Journal of Nuclear Materials},
  volume  = {406},
  number  = {1},
  pages   = {19--38},
  year    = {2010}
}

@article{agnew2005plastic,
  author  = {Agnew, S. R. and Duygulu, {\"O}.},
  title   = {Plastic Anisotropy and the Role of Non-Basal Slip in Magnesium Alloy {AZ31B}},
  journal = {International Journal of Plasticity},
  volume  = {21},
  number  = {6},
  pages   = {1161--1193},
  year    = {2005}
}

@article{wu2018mechanistic,
  author  = {Wu, Zhaoxuan and Ahmad, Rasool and Yin, Binglun and Sandl{\"o}bes, Stefanie and Curtin, W. A.},
  title   = {Mechanistic Origin and Prediction of Enhanced Ductility in Magnesium Alloys},
  journal = {Science},
  volume  = {359},
  number  = {6374},
  pages   = {447--452},
  year    = {2018}
}

@article{pollock2010weight,
  author  = {Pollock, Tresa M.},
  title   = {Weight Loss with Magnesium Alloys},
  journal = {Science},
  volume  = {328},
  number  = {5981},
  pages   = {986--987},
  year    = {2010}
}

@article{barnett2007twinning,
  author  = {Barnett, M. R.},
  title   = {Twinning and the Ductility of Magnesium Alloys: Part I. Tension Twins},
  journal = {Materials Science and Engineering: A},
  volume  = {464},
  number  = {1-2},
  pages   = {1--7},
  year    = {2007}
}

@article{fey2019pyg,
  author  = {Fey, Matthias and Lenssen, Jan Eric},
  title   = {Fast Graph Representation Learning with {PyTorch Geometric}},
  journal = {ICLR Workshop on Representation Learning on Graphs and Manifolds},
  year    = {2019}
}

@inproceedings{paszke2019pytorch,
  author    = {Paszke, Adam and Gross, Sam and Massa, Francisco and Lerer, Adam and Bradbury, James and Chanan, Gregory and Killeen, Trevor and Lin, Zeming and Gimelshein, Natalia and Antiga, Luca and others},
  title     = {{PyTorch}: An Imperative Style, High-Performance Deep Learning Library},
  booktitle = {Advances in Neural Information Processing Systems (NeurIPS)},
  volume    = {32},
  year      = {2019}
}

@inproceedings{loshchilov2019adamw,
  author    = {Loshchilov, Ilya and Hutter, Frank},
  title     = {Decoupled Weight Decay Regularization},
  booktitle = {International Conference on Learning Representations (ICLR)},
  year      = {2019}
}

@article{hendrycks2016gelu,
  author  = {Hendrycks, Dan and Gimpel, Kevin},
  title   = {Gaussian Error Linear Units ({GELU}s)},
  journal = {arXiv preprint arXiv:1606.08415},
  year    = {2016}
}

@article{ba2016layernorm,
  author  = {Ba, Jimmy Lei and Kiros, Jamie Ryan and Hinton, Geoffrey E.},
  title   = {Layer Normalization},
  journal = {arXiv preprint arXiv:1607.06450},
  year    = {2016}
}

@inproceedings{bengio2009curriculum,
  author    = {Bengio, Yoshua and Louradour, J{\'e}r{\^o}me and Collobert, Ronan and Weston, Jason},
  title     = {Curriculum Learning},
  booktitle = {Proceedings of the 26th Annual International Conference on Machine Learning (ICML)},
  pages     = {41--48},
  year      = {2009}
}

@article{mahata2025gnnal,
  author  = {Immanuel, Judah and Maiti, Aniruddha and Mahata, Avik},
  title   = {Graph Neural Network Surrogates for Direct Position Prediction in Molecular Dynamics of Face-Centered Cubic Aluminum},
  journal = {Manuscript in preparation},
  year    = {2025}
}

@article{MAHATA2026114443,
title = {Development and validation of interatomic potential for Sc and Al--Sc alloys: Thermodynamics, solidification, and intermetallic ordering},
journal = {Computational Materials Science},
volume = {264},
pages = {114443},
year = {2026},
issn = {0927-0256},
doi = {https://doi.org/10.1016/j.commatsci.2025.114443},
url = {https://www.sciencedirect.com/science/article/pii/S0927025625007864},
author = {Avik Mahata}
}

@article{MAHATA2022110902,
title = {Modified embedded-atom method interatomic potentials for Al-Cu, Al-Fe and Al-Ni binary alloys: From room temperature to melting point},
journal = {Computational Materials Science},
volume = {201},
pages = {110902},
year = {2022},
issn = {0927-0256},
doi = {https://doi.org/10.1016/j.commatsci.2021.110902},
url = {https://www.sciencedirect.com/science/article/pii/S0927025621006108},
author = {Avik Mahata and Tanmoy Mukhopadhyay and Mohsen {Asle Zaeem}}
}

@article{MAHATA2019125255,
title = {Effects of solidification defects on nanoscale mechanical properties of rapid directionally solidified Al-Cu Alloy: A large scale molecular dynamics study},
journal = {Journal of Crystal Growth},
volume = {527},
pages = {125255},
year = {2019},
issn = {0022-0248},
doi = {https://doi.org/10.1016/j.jcrysgro.2019.125255},
url = {https://www.sciencedirect.com/science/article/pii/S0022024819304701},
author = {Avik Mahata and Mohsen {Asle Zaeem}}
}

@article{Mahata_2018,
doi = {10.1088/1361-651X/aa9f36},
url = {https://doi.org/10.1088/1361-651X/aa9f36},
year = {2018},
month = {jan},
publisher = {IOP Publishing},
volume = {26},
number = {2},
pages = {025007},
author = {Mahata, Avik and Zaeem, Mohsen Asle and Baskes, Michael I},
title = {Understanding homogeneous nucleation in solidification of aluminum by molecular dynamics simulations},
journal = {Modelling and Simulation in Materials Science and Engineering}
}

@article{Mendelev2003,
  author    = {Mendelev, M. I. and Han, S. and Srolovitz, D. J. and Ackland, G. J. and Sun, D. Y. and Asta, M.},
  title     = {Development of New Interatomic Potentials Appropriate for Crystalline and Liquid Iron},
  journal   = {Philosophical Magazine},
  year      = {2003},
  volume    = {83},
  number    = {35},
  pages     = {3977--3994},
  doi       = {10.1080/14786430310001613264}
}

@article{Zhou2004,
  author    = {Zhou, X. W. and Johnson, R. A. and Wadley, H. N. G.},
  title     = {Misfit-Energy-Increasing Dislocations in Vapor-Deposited CoFe/NiFe Multilayers},
  journal   = {Physical Review B},
  year      = {2004},
  volume    = {69},
  number    = {14},
  pages     = {144113},
  doi       = {10.1103/PhysRevB.69.144113}
}

\end{document}